\documentclass[12pt]{iopart}
\usepackage{graphicx,amsfonts,amssymb,array,calc,tabularx,dsfont,amsthm,bm,mathrsfs,color}
\usepackage{bm}
\usepackage{supertabular}
\usepackage{booktabs}
\newcommand\tor{\mathds{T}^3}
\newcommand\Lie{\mathcal{L}}

\newcommand\lpl{l_{\mathrm{Pl}}}
\newcommand\Cyl{\mathrm{Cyl}_S}

\newcommand\HS{\mathcal{H}^S}
\newcommand\ket{|\mathbf{\vec\mu,\vec k}\rangle}

\newtheorem{thm}{Theorem}
\newtheorem{Def}{Definition}
\newtheorem{Prop}{Proposition}

\def\be{\begin{equation}}
\def\ee{\end{equation}}
\def\ba{\begin{eqnarray}}
\def\ea{\end{eqnarray}}
\def\ben{\begin{equation*}}
\def\een{\end{equation*}}
\def\ban{\begin{eqnarray*}}
\def\ean{\end{eqnarray*}}

\begin{document}

\title[LQC on a Torus]{Loop Quantum Cosmology on a Torus}
\author{Raphael Lamon}
\address{Institut f\"ur Theoretische Physik, Universit\"at Ulm \\ Albert-Einstein-Allee 11 \\ D-89069 Ulm, Germany}
\ead{raphael.lamon@uni-ulm.de}

\begin{abstract}
In this paper we study the effect of a torus topology on Loop Quantum Cosmology. We first derive the Teichm\"uller space parametrizing all possible tori using Thurston's theorem and construct a Hamiltonian describing the dynamics of these torus universes. We then compute the Ashtekar variables for a slightly simplified torus such that the Gauss constraint can be solved easily. We perform a canonical transformation so that the holomies along the edges of the torus reduce to a product between almost and strictly periodic functions of the new variables. The drawback of this transformation is that the components of the densitized triad become complicated functions of these variables. Nevertheless we find two ways of quantizing these components, which in both cases leads surprisingly to a continuous spectrum.
\index{\footnote{}}
\end{abstract}
\maketitle

\section{Introduction}\label{sec:intro}

The Einstein field equations are local equations in the sense that they only describe the local geometry of the spacetime. For example the Robertson-Walker metric explicitly contains the parameter $k$ which gives an account of the intrinsic spatial curvature. Using the Friedmann equations this parameter can be determined experimentally since it is directly related to the density parameter $\Omega_{\mathrm{tot}}$ and the Hubble parameter $h$. Recent measurements of the energy density of the universe tend to slightly favor a positively curved universe \cite{Komatsu:08}, yet a flat curvature lies within the 1-$\sigma$ range. The most direct conclusion is that the spatial topology of the universe is just $\mathds{R}^3$ which is the assumption of the $\Lambda$CDM model. Nevertheless in the mathematical literature it is well known that a flat space does not mean that its topology is necessarily $\mathds{R}^3$, in fact there are 18 possible flat topologies. Since the Einstein field equations are not sensitive to topology every possibility has to be considered as a possible candidate for the global geometry of our universe until it is ruled out by experiment. In order to do so we first note that the spectrum of the Laplace operator sensitively depends on the topology, i.e. it is discrete if the eigenstates are normalizable and continuous if not. In the first case the solution for e.g. a torus is given by plane waves with a wave vector $\vec k_n$ taking only discrete values $n\in\mathds{N}$ while in the second case  the (weak) solution to the eigenvalue equation is given by the (distributional) plane waves with a continuous wave vector $\vec k$. For example, the eigenvalue problem for $\triangle$ on $\mathds{T}^3$ is given by
$(\triangle+E_{\vec n})\Psi_{\vec n}=0,\; \vec n\in\mathds{N}^3$,
and on $S^3$ by
$\triangle\Psi_{\beta,l,m}=(\beta^2-1)\Psi_{\beta,l,m},$
where $\beta\in\mathds{N}$, $0\leq l\leq\beta-1$ and $|m|\leq l$. The implication of a solution of the form $\Psi_{\vec n}$ is the existence of a wave function $\Psi_n$ with a maximum length corresponding to e.g. the length of the edges of the torus. Since the departure from a continuous solution is biggest for large wavelengths we have to look for large-scale structures of the universe in order to distinguish between cosmic topologies. The best way to do so is to measure the inhomogeneities of the cosmic microwave background (CMB), expand these in multipole moments and compare the low multipoles with the predictions from theory. It can be shown that in certain closed topologies a suppression in the power spectrum of the low multipoles is expected because of the existence of a largest wavelength. Since such a suppression is present in the CMB several studies compared the theoretical predictions for various topologies with the data. While most analyzed topologies can already be ruled out three of them describe the data even better than the infinite $\Lambda$CDM model, namely the torus \cite{Aurich:08,Aurich:08:2,Aurich:09}, the dodecahedron\cite{Aurich:05,Caillerie:07} and the binary octahedron\cite{Aurich:05:2} (see also references therein). While the last two topologies are spherical the torus is the simplest model of a closed flat topology.

However, we know that standard cosmology cannot be the final answer as its predictability breaks down at the big bang. A quantization of the Friedmann equations a la Wheeler-DeWitt does not improve this behavior either. This situation has changed thanks to a new model called loop quantum cosmology (LQC) developed over the last few years which removes the initial singularity. LQC \cite{Bojowald:08,Bojowald:00,Bojowald:00:2,Bojowald:00:3,Bojowald:02,Ashtekar:03,Bojowald:03,Ashtekar:06,Ashtekar:06:2} is the approach motivated by loop quantum gravity (LQG) \cite{Ashtekar:04,Rovelli:04,Thiemann:07} to the quantization of symmetric cosmological models. The usual procedure is to reduce the classical phase space of the full theory to a  phase space with a finite number of degrees of freedom. The quantization of these reduced models uses the tools of the LQG and is therefore called LQC but it does not correspond to the cosmological sector of LQG. The results of LQC not only provide new insights into the quantum structure of spacetime near the Big-Band singularity but also remove this singularity by extending the time evolution to negative times.

In sum, on the one hand we have hints from observation that our universe may have a closed topology, on the other hand we have a very successful loop quantization of various cosmologies. Thus, starting from these two motivations, we would like to study LQC with a torus topology. But contrary to the works on the CMB we don't want to restrict the analysis to a cubical torus. To do so we construct a torus using Thurston's theorem and find that the most general torus has six degrees of freedom which consist of e.g. three lengths and three angles. We will study its dynamics by numerically solving the Hamiltonian coupled to a scalar field. After rewritting this Hamiltonian in terms of Ashtekar variables we will see that the quantization of such a torus leads to a product between the standard Hilbert spaces of LQC and the Hilbert spaces over the circle. Moreover, we will find two ways to quantize the components of the triad and show that both (generalized) eigenfunctions are not normalizable in this Hilbert space.

As a side remark we would like to point out that the consequences of putting a non-abelian gauge theory into a box with periodic boundary conditions have been studied in e.g. \cite{thooft:79}. The motivation behind this idea is an attempt to explain the quark confinement in QCD without explicitely breaking gauge invariance. To simplify the analysis the $su(N)$-valued gauge field is chosen to be pure gauge, i.e. $A=U^{-1}dU$ with $U\in SU(N)$, such that the holonomy around a closed curve $C$ only depends on the topological property of $C$. Since general relativity written in terms of Ashtekar variables is also a (constrained) Yang-Mills theory it may be tentalizing to use the methods developed for QCD in a box to LQC of a torus universe. However we will derive an Ashtekar connection for the homogeneous torus which is not pure gauge so that the holonomies along $C$ also depend on the length of $C$. This may not be surprising in view of the fact that the Hilbert space of LQC on $\mathds{R}^3$ is spanned by almost periodic functions with an arbitrary length parameter $\mu$.

This paper is organized as follows: in \Sref{sec:CHU} we first introduce the classical dynamics of a torus universe and numerically solve the Friedmann equations with a massless scalar field. In \Sref{sec:symmetryreduction} we introduce the Ashtekar variables for a torus and also explain the complications that arise because of a closed topology. The loop quantization and the construction of a Hilbert space are explained in \Sref{sec:Hkin} and \Sref{sec:conclusions} provides a summary and directions for future works. \ref{sec:fundamentaldomain} gives a short review of the fundamental domain of the 3-torus and \ref{sec:TorusIwasawa} describes the dynamics of the torus in terms of Iwasawa coordinates.

\section{Compact Homogeneous Universes and their Dynamics}\label{sec:CHU}
The purpose of this section is to study models in which the spatial section has a compact topology. The compactness of a locally homogeneous space brings new degrees of freedom of deformations, known as Teichm\"uller deformations. This leads to the conclusion that cosmology on a torus is simply cosmology on $\mathds{R}^3$ restricted to a cube may be too naive a point of view, especially since the space of solutions of a torus gets nine additional degrees of freedom, as already mentioned in \cite{Ashtekar:91}. We will introduce Teichm\"uller spaces with an emphasis on a Thurston geometry admitting a Bianchi I geometry as its subgeometry \cite{Wolf:74,Koike:94,Tanimoto:97,Tanimoto:97:02,Yasuno:01} and derive the vacuum Friedmann equations using the Hamiltonian formalism.

\subsection{Compact Homogeneous Spaces}\label{sec:CHS}
Let $\Sigma$ be a three-dimensional, arcwise connected Riemannian manifold.
\begin{Def} 
A metric on a manifold $\Sigma$ is locally homogeneous if $\forall p,q\in \Sigma$ there exist neighborhoods $U,V$ of $p$ resp. $q$ and an isometry $(U,p)\rightarrow (V,q)$. The manifold is globally homogeneous if the isometry group acts transitively on the whole manifold $\Sigma$.
\end{Def}
Since $\Sigma$ is arcwise connected we know that there is a unique universal covering manifold $\tilde \Sigma$ up to diffeomorphisms with a metric given by the pullback of the metric on $\Sigma$ by the covering map 
\begin{equation}\label{coveringmap}
\pi:\tilde \Sigma\rightarrow \Sigma. 
\end{equation}
Singer \cite{Singer:60} proved that the metric on $\tilde \Sigma$ is then globally homogeneous and $\tilde \Sigma$ is given by $\tilde \Sigma\cong\tilde S/F$, where $\tilde S$ is the orientation preserving isometry group of $\tilde \Sigma$ and $F$ its isotropy subgroup.

On the other hand, we can also start from a three-dimensional, simply connected Riemannian manifold $\tilde \Sigma$ which admits a compact quotient $\Sigma$. In order to construct this compact manifold consider the covering group $\Gamma\subset \tilde S$ which is isomorphic to the fundamental group $\pi_1(\Sigma)$ of $\Sigma$. This implies that
\begin{equation*}
\Sigma=\tilde \Sigma/\Gamma,
\end{equation*}
which is Hausdorff iff $\Gamma$ is a discrete subgroup of $\tilde S$ and a Riemannian manifold iff $\Gamma$ acts freely on $\tilde \Sigma$.
\begin{Def}
A geometry is the pair $(\tilde\Sigma,S)$ where $\tilde S$ a group acting transitively on $\tilde \Sigma$ with compact isotropy subgroup. A geometry $(\tilde \Sigma,\tilde S')$ is a subgeometry of $(\tilde \Sigma,\tilde S)$ if $\tilde S'$ is a subgroup of $\tilde S$. A geometry $(\tilde \Sigma,\tilde S)$ is called maximal if it is not a subgeometry of any geometry and minimal if it does not have any subgeometry.
\end{Def}
We will need the following important theorem:
\begin{thm}[Thurston \cite{Thurston:97}] Any maximal, simply connected 3-dimensional geometry which admits a compact quotient is equivalent to the geometry $(\tilde \Sigma,\tilde S)$ where $\tilde \Sigma$ is one of $E^3$ (Euclidean), $H^3$ (hyperbolic), $S_p^3$ (3-sphere), $S_p^2\times\mathds{R}$, $H^2\times\mathds{R}$, $\widetilde{SL}(2,\mathds{R})$, Nil or Sol.
\end{thm}
If $(\tilde \Sigma,\tilde S')$ is not a maximal geometry but is simply connected and admits a compact quotient as well we can find a discrete subgroub $\Gamma'$ of $\tilde S'$ acting freely so as to make $\tilde \Sigma/\Gamma'$ compact. Define $(\tilde \Sigma,\tilde S)$ as the maximal geometry with $(\tilde \Sigma,\tilde S')$ as its subgeometry, i.e. $\tilde S'\subset \tilde S$. By Thurston's Theorem $(\tilde \Sigma,\tilde S)$ is one of the eight Thurston geometries, which implies that $(\tilde \Sigma,\tilde S')$ is a subgeometry of one of the eight Thurston geometries.
\begin{thm} Any minimal, simply connected three-dimensional geometry is equivalent to $(\tilde \Sigma,\tilde S)$, where $\tilde \Sigma=\mathds{R}^3$, $\tilde S=$Bianchi I-VIII; $\tilde \Sigma=S_p^3$, $\tilde S=$Bianchi IX; or $\tilde \Sigma=S_p^2\times\mathds{R}$, $\tilde S=SO(3)\times\mathds{R}$, where $S_p^3$ is the three-sphere and $S_p^2$ the two-sphere.
\end{thm}
Let Rep$(\Sigma)$ denote the space of all discrete and faithful representations $\rho:\pi_1(\Sigma)\rightarrow \tilde S$ and the diffeomorphism $\phi:\tilde \Sigma\rightarrow \tilde \Sigma$ a {\it global conformal isometry} if $\phi_*\tilde h_{ab}=\mathrm{const}\cdot \tilde h_{ab}$, where $\tilde h_{ab}$ is the spatial metric of the universal covering manifold $\tilde \Sigma$. This allows us to define a relation $\rho\sim\rho'$ in Rep$(\Sigma)$ if there exists a conformal isometry $\phi$ of $\tilde \Sigma$ connected to the identity with $\rho'(a)=\phi\circ\rho(a)\circ\phi'$.

\begin{Def}\label{def:Teichmueller}
The Teichm\"uller space is defined as
\begin{equation*}
\mathrm{Teich}(\Sigma)=\mathrm{Rep}(\Sigma)/\sim
\end{equation*}
with elements called Teichm\"uller deformations, which are smooth and nonisometric deformations of the spatial metric $h_{ab}$ of $\Sigma$, leaving the universal cover $(\tilde \Sigma,\tilde h_{ab})$ globally conformally isometric.
\end{Def}

The situation gets more complicated when we try to extend the previous construction to four-dimensional Lorentzian manifolds. The reason is that the action of the covering group $\Gamma$ needs to preserve both the extrinsic curvature and the spatial metric of $\tilde \Sigma$. Thus we cannot construct a homogeneous compact manifold by the action of a discrete subgroup of $\tilde S$ on the spatial three-section $\tilde \Sigma$. Instead we need the isometry group of the four-dimensional manifold $\tilde M$. Let $M=\mathds{R}\times \Sigma$ be a compact homogeneous Lorentzian manifold with metric $g_{\mu\nu}$ and $\tilde M=\mathds{R}\times\tilde\Sigma$ its covering with metric $\tilde g_{\mu\nu}$  $(\mu,\nu=0,\ldots,4)$.
\begin{Def} Let $(\tilde \Sigma,\tilde h_{ab})$ be a spatial section of $(\tilde M,\tilde g_{\mu\nu})$. An extendible isometry is defined by the restriction of an isometry of $(\tilde M,\tilde g_{ab})$ on $\tilde \Sigma$ which preserves $\tilde \Sigma$ and forms a subgroup $\mathrm{Esom}(\tilde \Sigma)$ of $\tilde S$.
\end{Def}
Thus, in order to get a compact homogeneous manifold from $\tilde M$ the covering group $\Gamma$ must be a subgroup of Esom$(\tilde \Sigma)$, i.e.
\begin{equation*}
\Gamma\subset\mathrm{Esom}(\tilde \Sigma).
\end{equation*}
The line element of $M=\mathds{R}\times \Sigma$ is given by
$$ds^2=-dt^2+h_{ab}(t)\sigma^a\sigma^b,$$
where $\sigma^a$ are the invariant one-forms.

Therefore the Teichm\"uller parameters enlarge the parameter space by bringing new degrees of freedom from the deformations defined in Definition~\ref{def:Teichmueller}. In fact, the set of all possible universal covers $(\tilde M,\tilde g_{ab})$ carries the degrees of freedom of the local geometry and the covering maps $\Gamma$ the degrees of freedom of the global geometry which are parameterized by the Teichm\"uller parameters.

\subsection{The Torus Universe}\label{sec:torusuniverse}
In this section we restrict the above analysis to the case of a flat torus and give only the main results. Further details can be found in \cite{Wolf:74,Koike:94,Tanimoto:97,Tanimoto:97:02,Yasuno:01}. Let $\tilde M=\mathds{R}\times\tilde \Sigma$ be the universal cover of $M$ and $\tilde \Sigma$ the Thurston geometry $(E^3,ISO(3,\mathds{R}))$. The isometry group $ISO(3)$ is expressed as $g(\mathbf{x})=\mathbf{Rx}+\mathbf{a}$, where $\mathbf{a}$ is a constant vector and $\mathbf{R}\in SO(3)$ in order that the orientation be preserved. The Killing vectors of $E^3$ are
\begin{eqnarray*}
\xi_1=\partial_x,\quad \xi_2=\partial_y,\quad \xi_3=\partial_z,\nonumber \\
\xi_4=-z\partial_y+y\partial_z,\quad \xi_5=-x\partial_z+z\partial_x,\quad \xi_6=-y\partial_x+x\partial_y.
\end{eqnarray*}
The line element of $\tilde M$ is thus given by
$$ds^2=-dt^2+\tilde h_{ab}dx^adx^b=-dt^2+a^2(t)\,^0\tilde h_{ab}dx^adx^b,$$
where $\,^0\tilde h_{ab}$ is called the fiducial metric in the LQC literature and $dx^a$ are the invariant one-forms of the group $ISO(3,\mathds{R})$ \footnote{When dealing with the open case $\mathds{R}^3$ one has to distinguish between the fiducial volume $V_0$ of a cell as measured by the fiducial metric $\,^0\tilde h_{ab}$ and the physical volume $V$ as measured by the physical metric $\tilde h_{ab}$. Since we shall deal with a closed universe we have the "preferred fiducial cell" $\mathds{T}^3$ at our disposal. Furthermore, in the open case the spatial integrals have to be restricted to this fiducial cell whereas in the closed case these integrals are naturally restricted to the physical cell $\mathds{T}^3$.}. 

\begin{figure}[!ht]
 \begin{center}
 \includegraphics[width=8cm]{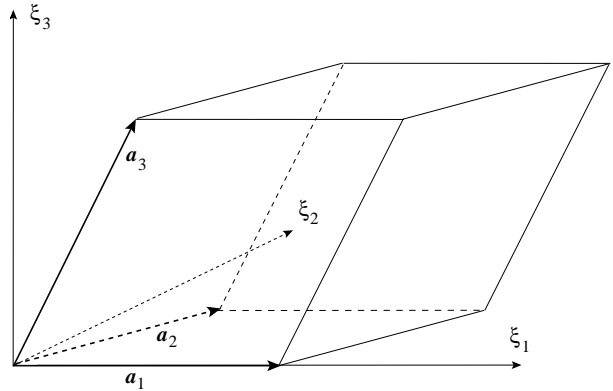}
\caption{The vectors $a_1$, $a_2$ and $a_3$ span the torus with six Teichm\"uller parameters. The global conformal invariance was used in order to align $a_1$ with $\xi_1$ and $a_2$ with span$\{\xi_1,\xi_2\}$.}\label{fig:torus}
\end{center}
\end{figure}

The covering group $\Gamma\subset\mathrm{Esom}(\tilde\Sigma)\equiv \mathrm{Esom}(E^3)$ allows us to construct a torus via $M=\tilde M/\Gamma$, where $M=\mathds{R}\times\mathds{T}^3$. The freedom of global conformal transformations allows us to choose the coordinate system of $\tilde \Sigma$ such that the generators of the torus have a simple representation. We thus require one of the generators to be aligned with $\xi_1$ and one to lie in the $\xi_1\xi_2$-plane.  The Teichm\"uller space is then generated by six Teichm\"uller parameters in three vectors
\begin{equation}\label{Teichmuellervectors}
a_1=\left(
\begin{array}{c}
a_1\,^1 \\ 0 \\ 0
\end{array}
\right),
\quad
a_2=\left(
\begin{array}{c}
a_2\,^1 \\ a_2\,^2 \\ 0
\end{array}
\right),
\quad
a_3=\left(
\begin{array}{c}
a_3\,^1 \\ a_3\,^2 \\ a_3\,^3
\end{array}
\right),
\end{equation}
where all $a_i\,^j$ only depend on the coordinate time $t$.
The configuration space $\mathcal{C}$ is therefore spanned by the six Teichm\"uller parameters such that $\mathcal{C}\subset\mathds{R}^6$ (see \ref{sec:fundamentaldomain}). The {\it flat} spatial metric on $\tor$ is then given by $(a,b=1,2,3)$
\begin{equation}\label{ds2Teich}
ds^2=h_{ab}dx^{a}dx^{b},\quad h_{ab}=\sum_c a_a\,^ca_b\,^c
\end{equation}
where
\begin{equation}\label{metrich}
(h_{ab})=\left(
\begin{array}{ccc}
(a_1\,^1)^2 & a_1\,^1a_2\,^1 & a_1\,^1a_3\,^1 \\
& (a_2\,^1)^2+(a_2\,^2)^2 & a_2\,^1a_3\,^1+a_2\,^2a_3\,^2 \\
\mathrm{(sym)} & & (a_3\,^1)^2+(a_3\,^2)^2+(a_3\,^3)^2
\end{array}
\right).
\end{equation}
This metric is invariant under transformations in $SL(3,\mathds{Z})$. For example it is left invariant by $(a_1\,^1\rightarrow -a_1\,^1,a_2\,^1\rightarrow -a_2\,^1,a_3\,^1\rightarrow -a_3\,^1)$ (see \ref{sec:fundamentaldomain} for more details). From \Eref{metrich} we can make a Legendre transform of the Einstein-Hilbert action 
\begin{equation}\label{SEH}
S_{\mathrm{E-H}}[g]=\frac{1}{2\kappa}\int_{\mathds{R}\times\tor}*R[g], \quad \kappa=8\pi G
\end{equation}
to obtain a Hamiltonian, where $*$ is the Hodge star operator. After a partial integration of $\ddot a_i\,^i$ (which also cancels the surface term we omitted in \Eref{SEH}) we find the following Lagrangian:
\ban
\mathcal{L}=&&\frac{1}{4\kappa}\frac{1}{a_1\,^1a_2\,^2a_3\,^3}\times\\
&\biggl[& \left((a_3\,^2)^2+(a_3\,^3)^2\right)\left(a_2\,^1\dot a_1\,^1-a_1\,^1\dot a_2\,^1\right)^2+(a_1\,^1)^2(a_3\,^2)^2(\dot a_2\,^2)^2\\
&+&(a_2\,^2)^2\Bigl\{(a_3\,^1)^2(\dot a_1\,^1)^2-2a_1\,^1a_3\,^1\dot a_1\,^1\dot a_3\,^1\\
&&\quad\quad\quad+a_1\,^1\Bigl(a_1\,^1\left((\dot a_3\,^1)^2+(\dot a_3\,^2)^2\right)-4a_3\,^3\dot a_1\,^1\dot a_3\,^3\Bigr)\Bigr\}\\
&-&2a_2\,^2\Bigl\{a_2\,^1a_3\,^2\dot a_1\,^1\left(a_3\,^1\dot a_1\,^1-a_1\,^1\dot a_3\,^1\right)\\
&&\quad\quad\quad+a_1\,^1\Bigl[a_1\,^1a_3\,^2\dot a_2\,^1\dot a_3\,^1-a_3\,^1a_3\,^2\dot a_1\,^1\dot a_2\,^1\\
&&\quad\quad\quad\quad\quad\quad+\dot a_2\,^2\left(a_1\,^1a_3\,^2\dot a_3\,^2+2a_3\,^3(a_3\,^3\dot a_1\,^1+a_1\,^1\dot a_3\,^3)\right)\Bigr]\Bigr\}\biggl]
\ean
We introduce the momenta 
\begin{equation}\label{defp}
p^a\,_b:=\frac{\partial\mathcal{L}}{\partial \dot a_a\,^b}
\end{equation}
conjugate to the configuration variables $a_a\,^b$ such that the phase space $\mathcal{P}=T^*\mathcal{C}\subset\mathds{R}^{12}$ is the cotangent bundle over $\mathcal{C}$ with
\begin{equation}\label{bracketap}
\{a_a\,^b,p^c\,_d\}=\delta_a^c\delta^b_d, \quad \{a_a\,^b,a_c\,^d\}=0, \quad \{p^a\,_b,p^c\,_d\}=0,
\end{equation}
where the Poisson brackets are defined as
$$\{f,g\}=\sum_{a,b=1}^{3}\frac{\partial f}{\partial a_a\,^b}\frac{\partial g}{\partial p^a\,_b}-\frac{\partial g}{\partial a_a\,^b}\frac{\partial f}{\partial p^a\,_b}$$
for any smooth functions on the phase space. We insert $\dot a_a\,^b=\dot a_a\,^b(p^c\,_d)$ into the Legendre transform of \Eref{SEH} and get the Hamiltonian
\ba\label{Hamiltonian}
\mathcal{H}_g=&&\frac{\kappa}{4}\frac{1}{a_1\,^1a_2\,^2a_3\,^3} \times \nonumber \\
&\biggl[&(a_1\,^1p^1\,_1)^2+(a_2\,^2p^2\,_2)^2+(a_3\,^3p^3\,_3)^2+(a_2\,^1p^2\,_1)^2+4(a_2\,^2p^2\,_1)^2\nonumber\\
&+&(a_3\,^1p^3\,_1)^2+4(a_3\,^2p^3\,_1)^2+4(a_3\,^3p^3\,_1)^2+(a_3\,^2p^3\,_2)^2\nonumber\\
&+&4(a_3\,^3p^3\,_2)^2-2a_3\,^2a_3\,^3p^3\,_2p^3\,_3\nonumber\\
&+&2a_1\,^1p^1\,_1\left(a_2\,^1p^2\,_1-a_2\,^2p^2\,_2+a_3\,^1p^3\,_1-a_3\,^2p^3\,_2-a_3\,^3p^3\,_3\right)\\
&-&2a_3\,^1p^3\,_1\left(a_3\,^2p^3\,_2+a_3\,^3p^3\,_3\right)\nonumber\\
&-&2a_2\,^1p^2\,_1\left(a_2\,^2p^2\,_2-a_3\,^1p^3\,_1+a_3\,^2p^3\,_2+a_3\,^3p^3\,_3\right)\nonumber\\
&+&2a_2\,^2\Bigl\{a_3\,^2\left(4p^2\,_1p^3\,_1+p^2\,_2p^3\,_2\right)-p^2\,_2\left(a_3\,^1p^3\,_1+a_3\,^3p^3\,_3\right)\Bigr\}\biggr]\nonumber
\ea
The Hamiltonian constraint $\mathcal{H}_g\approx0$ reduces the dynamical degrees of freedom from dim~$\mathcal{P}=12$ to dim~$\mathcal{P}=10$, which agrees with \cite{Ashtekar:91}. To compare this Hamiltonian with the usual Bianchi type I models we set all offdiagonal elements to zero and $a_i\,^i=a_i$, $p^i\,_i=p^i$ (no summation), and get
\be
\mathcal{H}_g=\frac{\kappa}{4}\left(\frac{a_1(p^1)^2}{a_2a_3}+\frac{a_2(p^2)^2}{a_1a_3}+\frac{a_3(p^3)^2}{a_1a_2}-2\frac{p^1p^2}{a_3}-2\frac{p^2p^3}{a_1}-2\frac{p^1p^3}{a_2}\right),
\ee
which agrees with the result given in \cite{Chiou:07} up to a factor 2 in the definition of the action. To get the isotropic case\footnote{At this point care has to be taken because there is no homogeneous and isotropic vacuum solution to the Einstein equation (see \Sref{sec:openmodels}). Only a nonvanishing energy-momentum tensor allows for the isotropic limit of $\mathcal{H}_g$, which corresponds then to the usual Friedmann solutions.} we further set $a_i=a$, $p^i=p/3$ and find that the Hamiltonian \eref{Hamiltonian} reduces to the usual first Friedmann equation
\begin{equation*}
\mathcal{H}_g=-\frac{\kappa p^2}{12a}
\end{equation*}
and the Hamiltonian equation $\dot p^i\,_j=-\partial \mathcal{H}_g/\partial a_i\,^j$ to the usual second Friedmann equation
\begin{equation*}
\dot p=-\frac{\partial \mathcal{H}_g}{\partial a}=\frac{\kappa p^2}{12a^2}.
\end{equation*}
The second Hamiltonian equation is given by $\dot a=\partial \mathcal{H}_g/\partial p=-\kappa p/(6a)$ and allows us to recast the first Friedmann equation into the usual form 
$$\mathcal{H}_g=-3a\dot a^2/\kappa.$$
Furthermore, notice that all $a_i\,^j$ and $p^i\,_j$, $i\neq j$, have to vanish in order for the torus to remain aligned with the Killing fields $\xi_I$.

\begin{figure}[!ht]
 \begin{center}
 \includegraphics[width=7.4cm]{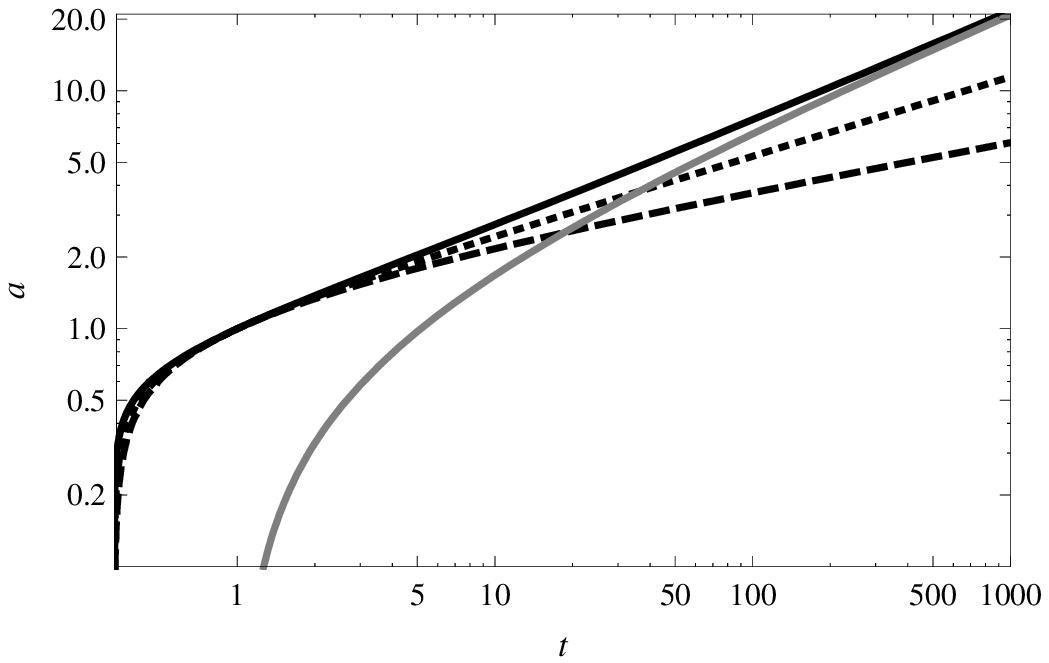}\includegraphics[width=7.9cm]{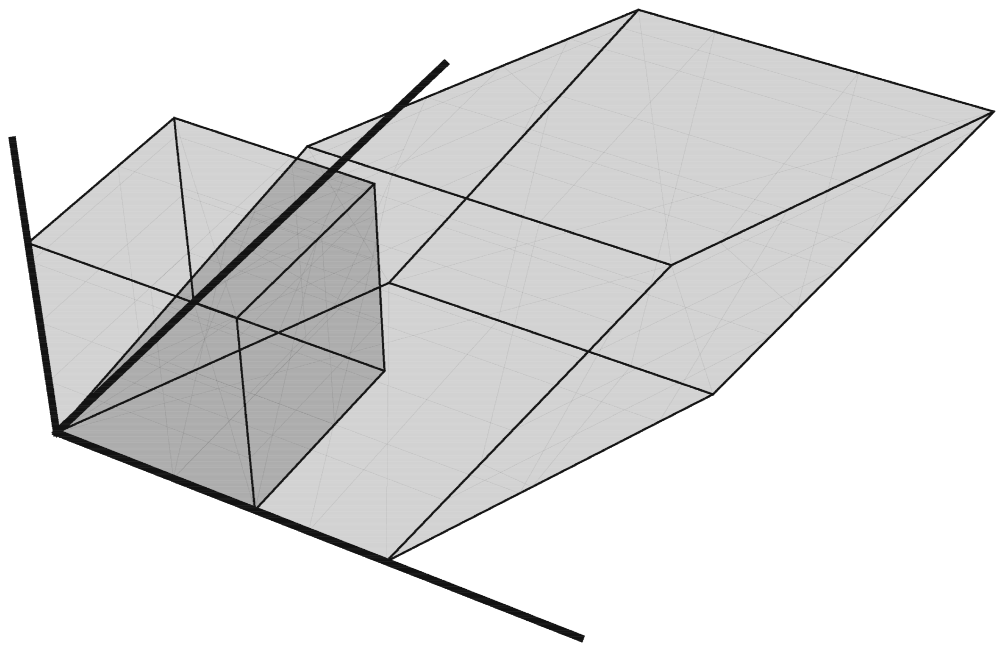}
\caption{{\it Left panel}: Solutions corresponding to the Hamiltonian \eref{Hamiltonianmatter} with the initial conditions $a_i\,^i(1)=1$, $p^i\,_i(1)=-1$ (no summation), $a_i\,^j(1)=0$ $(i\neq j)$, $p^3\,_1(1)=p^3\,_2(1)=0$, $p^2\,_1(1)=0.2$, $\phi(1)=10^{-3}$, $\pi(1)=1.2$. The diagonal momenta $p^i\,_i$ are chosen to be negative such that all sides of the torus expand.  The solid black line is $a_1\,^1$, the dashed one $a_2\,^2$, the dotted one $a_3\,^3$ and the gray one the off-diagonal $a_2\,^1$. The time $t$ parametrizes the coordinate time in natural units ($c=\kappa=\hbar=1$). {\it Right panel}: Solution corresponding to the Hamiltonian \eref{Hamiltonianmatter} at two different times. The initial condition is a cubic universe with $a_i\,^i\equiv a_0$, $p^i\,_i\equiv p_0$, $a_i\,^j=0$ $(i\neq j)$, $p^i\,_j\neq 0$ $(i\neq j)$. For both panels the mass and the potential of the scalar field have been set to zero.}\label{fig:torus_sim}
\end{center}
\end{figure}

We add a matter term consisting of a homogeneous massive scalar field\footnote{Notice that since every scalar field lives in the trivial representation of the rotation group it is not possible to construct a scalar field which is homogeneous but anisotropic.} to \Eref{Hamiltonian} to obtain the Hamiltonian
\begin{equation}\label{Hamiltonianmatter}
\mathcal{H}=\mathcal{H}_g+\mathcal{H}_{\phi}=\mathcal{H}_g+\frac{1}{2\sqrt{h}}\pi^2+\frac{\sqrt{h}}{2}m_{\phi}^2\phi^2+\sqrt{h}V(\phi),
\end{equation}
where $\pi$ is the momentum of the scalar field, $m_{\phi}$ its mass, $h=(a_1\,^1)^2(a_2\,^2)^2(a_3\,^3)^2$ the determinant of the spatial metric \eref{metrich} and $V(\phi)$ the potential which we set to zero in the sequel. From this equation we calculate the Friedmann equations and compute the shape of the universe for a special choice of initial conditions, which is shown in \Fref{fig:torus_sim}. All classical solutions have the limit $\lim_{t\rightarrow 0}a_i\,^j=0$ and grow with $a\propto t^{1/3}$ for a massless scalar field with zero potential. Furthermore, note the convergence of $a_1\,^1$ and $a_2\,^2$, which is explained in \ref{sec:TorusIwasawa}.

\section{Symmetry Reduction and Classical Phase Space for Ashtekar Variables}\label{sec:symmetryreduction}
In this section we shall repeat the complete analysis introduced in \cite{Kobayashi:63,Bojowald:00,Bojowald:00:2} in order to see the role of a compact topology on a connection. Our strategy is to find an invariant connection on the covering space $\tilde M$ and then restrict it to the compact space $M$ by means of the covering map \eref{coveringmap}. In the following section, when referring to the covering space, we shall use a tilde.

\subsection{Invariant Connections}\label{sec:invariantconnections}
Let $\tilde P(\tilde M,SU(2),\pi)$ be a principal fiber bundle over $\tilde M$  with structure group $SU(2)$ and projection $\pi:\tilde P\rightarrow \tilde M$. We require that there be a symmetry group $\tilde S\subset \mathrm{Aut}(\tilde P)$ of bundle automorphisms which acts transitively. Furthermore, for Bianchi I models $\tilde S$ does not have a non-trivial isotropy subgroup $\tilde F$ so that the base manifold is isomorphic to the symmetry group $\tilde S$, i.e. $\tilde M/\tilde S=\{x_0\}$ is represented by a single point that can be chosen arbitrarily in $\tilde M$. Since the isotropy group $\tilde F$ is trivial the coset space $\tilde S/\tilde F\cong \tilde S$ is reductive with a decomposition of the Lie algebra of $\tilde S$ according to $\Lie \tilde S=\Lie \tilde F\oplus\Lie \tilde F_{\perp}=\Lie \tilde F_{\perp}$ together with the trivial condition $\mathrm{Ad}_{\tilde F}\Lie \tilde F_{\perp}\subset\Lie \tilde F_{\perp}$. This allows us to use the general framework described in \cite{Bojowald:00,Bojowald:00:2,Kobayashi:63}.

Since the isotropy group plays an important role in classifying symmetric bundles and invariant connections we describe the general case of a general isotropy group $\tilde F$. Fixing a point $x\in\tilde M$, the action of $\tilde F$ yields a map $\tilde F:\pi^{-1}(x)\rightarrow\pi^{-1}(x)$ of the fiber over $x$. To each point $p\in\pi^{-1}(x)$ in the fiber we assign a group homomorphism $\lambda_p:\tilde F\rightarrow G$ defined by $f(p)=:p\cdot\lambda_p(f)$, $\forall f\in \tilde F$. As this homomorphism transforms by conjugation $\lambda_{p\cdot g}=\mathrm{Ad}_{g^{-1}}\circ\lambda_p$ only the conjugacy class $[\lambda]$ of a given homomorphism matters. In fact, it can be shown \cite{Kobayashi:63} that an $\tilde S$-symmetric principal bundle $P(\tilde M,G,\pi)$ with isotropy subgroup $\tilde F\subseteq \tilde S$ is uniquely characterized by a conjugacy class $[\lambda]$ of homomorphisms $\lambda:\tilde F\rightarrow G$ together with a reduced bundle $Q(\tilde M/\tilde S,Z_G(\lambda(\tilde F)),\pi_Q)$, where $Z_G(\lambda(\tilde F))$ is the centralizer of $\lambda(\tilde F)$ in $G$. In our case, since $\tilde F=\{1\}$ all homomorphisms $\lambda:\tilde F\rightarrow G=SU(2)$ are given by $1\mapsto 1_G$.

After having classified the $\tilde S$-symmetric fiber bundle $\tilde P$ we seek a $[\lambda]$-invariant connection on $\tilde P$. We use the following general result \cite{Brodbeck:96}:

\begin{thm}[Generalized Wang theorem] Let $\tilde P$ be an $\tilde S$-symmetric principal bundle classified by $([\lambda],Q)$ and let $\tilde \omega$ be a connection in $\tilde P$ which is invariant under the action of $\tilde S$. Then $\tilde \omega$ is classified by a connection $\tilde \omega_Q$ in $Q$ and a scalar field (usually called the Higgs field) $\phi:Q\times\Lie \tilde F_{\perp}\rightarrow\Lie G$ obeying the condition
\begin{equation}\label{condonphi}
\phi(\mathrm{Ad}_f(X))=\mathrm{Ad}_{\lambda(f)}\phi(X)$, $\forall f\in \tilde F,\; X\in\Lie  \tilde F_{\perp}.
\end{equation}
\end{thm}

The connection $\tilde \omega$ can be reconstructed from its classsifying structure as follows. According to the decomposition $\tilde M\cong\tilde M/\tilde S\times \tilde S/\tilde F$ we have $\tilde \omega=\tilde{\omega}_Q+\tilde \omega_{\tilde S/\tilde F}$ with $\tilde \omega_{\tilde S/\tilde F}=\phi\circ\iota^*\tilde \theta_{\mathrm{MC}}$, where $\iota:\tilde S/\tilde F\hookrightarrow \tilde S$ is a local embedding and $\tilde \theta_{\mathrm{MC}}$ is the Maurer-Cartan form on $\tilde S$. The structure group $G$ acts on $\phi$ by conjugation, whereas the solution space of \Eref{condonphi} is only invariant with respect to the reduced structure group $Z_G(\lambda(\tilde F))$. This fact leads to a partial gauge fixing since the connection form $\tilde \omega_{\tilde S/\tilde F}$ is a $Z_G(\lambda(\tilde F))$-connection which explicitly depends on $\lambda$. We then break down the structure group from $G$ to $Z_G(\lambda(\tilde F))$ by fixing a $\lambda\in[\lambda]$.

In our case, the embedding $\iota:\tilde S\rightarrow \tilde S$ is the identity and the base manifold $\tilde M/\tilde S=\{x_0\}$ of the orbit bundle is represented by a single point so that the invariant connection is given by
\begin{equation*}
\tilde A=\phi\circ\tilde \theta_{\mathrm{MC}}.
\end{equation*}
The three generators of $\Lie \tilde S$ are given by $T_I$, $1\leq I\leq 3$, with the relation $[T_I,T_J]=0$ for Bianchi I models. The Maurer-Cartan form is given by $\tilde \theta_{\mathrm{MC}}=\tilde \omega^I T_I$ where $\tilde \omega^I$ are the left invariant one-forms on $\tilde S$. The condition \eref{condonphi} is empty so that the Higgs field is given by $\phi:\Lie \tilde S\rightarrow\Lie G,\;T_I\mapsto \phi(T_I)=:\phi_I\,^i\tau_i$, where the matrices $\tau_j=-i\sigma_j/2$, $1\leq j\leq 3$, generate $\Lie G$, where $\sigma_j$ are the standard Pauli matrices\footnote{We use the convention $\tau_i\tau_j=\frac{1}{2}\epsilon_{ij}\,^k\tau_k-\frac{1}{4}\delta_{ij}\mathds{1}_{2\times 2}$}. In summary the invariant connection is given by
\begin{equation}\label{Anondiag}
\tilde A=\phi_I\,^i\tau_id\tilde \omega^I.
\end{equation}

In order to restrict this invariant connection we define the invariant connection $A$ on $\mathds{T}^3$ with the pullback given by the covering map \eref{coveringmap}. The generators of the Teichm\"uller space (see \Eref{Teichmuellervectors}) allow us to write $A$ as:
\begin{equation}\label{defofphigen}
A_a^i:=\bar \phi_I\,^i\omega_a^I,\quad (\bar\phi_I\,^i)=\left(
\begin{array}{ccc}
\bar\phi_1\,^1 & \bar\phi_2\,^1 & \bar\phi_3\,^1 \\
0 & \bar\phi_2\,^2 & \bar\phi_3\,^2 \\
0 & 0 & \bar\phi_3\,^3
\end{array}
\right).
\end{equation}

\subsubsection{Simplified Model}\label{sec:simplmodel}
\begin{figure}[!ht]
 \begin{center}
 \includegraphics[width=6cm]{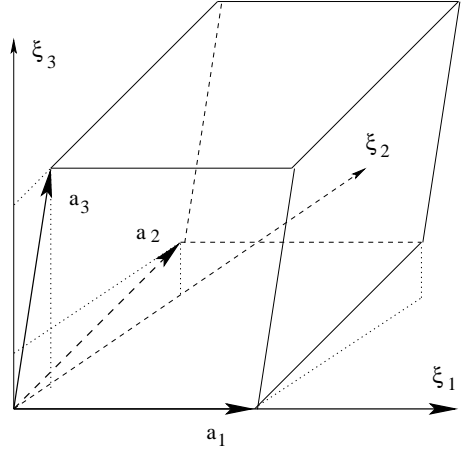}
\caption{The vectors $a_1$, $a_2$ and $a_3$ span the torus with five Teichm\"uller parameters. The vectors $a_2$ and $a_3$ lie in the $\xi_2\xi_3$-plane while $a_1$ is aligned with $\xi_1$}\label{fig:torus_simpl}
\end{center}
\end{figure}

In the sequel we shall concentrate on a simpler model for which we can also easily satisfy the Gauss constraint. We choose a torus generated by the vectors $a_1=(a_1\,^1,0,0)^T$, $a_2=(0,a_2\,^2,a_2\,^3)^T$ and $a_3=(0,a_3\,^2,a_3\,^3)^T$ (see \Fref{fig:torus_simpl}) such that
\be\label{defofphi} (\bar\phi_I\,^i)=\left(
\begin{array}{ccc}
\bar\phi_1\,^1 & 0 & 0 \\
0 & \bar\phi_2\,^2 & \bar\phi_3\,^2 \\
0 & \bar\phi_2\,^3 & \bar\phi_3\,^3
\end{array}
\right),\quad (\omega_a^I)=\left(
\begin{array}{ccc}
a_1\,^1 & 0 & 0 \\
0 & a_2\,^2 & a_3\,^2 \\
0 & a_2\,^3 & a_3\,^3
\end{array}
\right)
\ee 
and 
\ben (A_a^i)=\left(
\begin{array}{ccc}
a_1\,^1\bar\phi_1\,^1 & 0 & 0 \\
0 & a_2\,^2\bar\phi_2\,^2+a_2\,^3\bar\phi_3\,^2 & a_3\,^2\bar\phi_2\,^2+a_3\,^3\bar\phi_3\,^2 \\
0 & a_2\,^2\bar\phi_2\,^3+a_2\,^3\bar\phi_3\,^3 & a_3\,^2\bar\phi_2\,^3+a_3\,^3\bar\phi_3\,^3
\end{array}
\right)
\een
The vectors $X_I$ dual to $\omega^I$ are given by
\ben X_1=\left(
\begin{array}{c}
\frac{1}{a_1\,^1}\\ 0 \\ 0
\end{array}
\right), \;  
X_2=\frac{1}{\mathfrak{h}}\left(
\begin{array}{c}
0\\ a_3\,^3 \\ -a_2\,^3
\end{array}
\right), \;
X_3=\frac{1}{\mathfrak{h}}\left(
\begin{array}{c}
0 \\ -a_3\,^2 \\ a_2\,^2
\end{array}
\right), \een
where we defined $\mathfrak{h}=a_2\,^2a_3\,^3-a_2\,^3a_3\,^2$.

\subsection{Classical Phase Space for Ashtekar Variables}\label{sec:classicalphasespace}
The phase space of full general relativity in the Ashtekar representation is spanned by the $SU(2)$-connection $A_a^i=\Gamma_a^i+\gamma K_a^i$ and the densitized triad $E_i^a=|\det e|e_i^a$, where $\Gamma_a^i$ is the spin connection, $K_a^i$ the extrinsic curvature, $e_i^a$ the triad and $\gamma>0$ the Immirzi parameter \cite{Ashtekar:04,Rovelli:04,Thiemann:07}. The symplectic stucture of full general relativity is given by the Poisson bracket
\begin{equation}\label{AEvariables}
\{A^i_a(y),E^b_j(x)\}=\kappa\delta_a^b\delta_j^i\delta(x,y).
\end{equation}
The connection between the metric and the densitized triad is given by
\be\label{EEeqh} hh^{ab}=\delta^{ij}E_i^aE_j^b,\ee
where $h^{ab}$ is the inverse of the metric $h_{ab}$.

We can now use the results obtained in \Sref{sec:CHU} to construct the phase space $\mathcal{P}$ in this representation. In the preceding subsection we have already found the configuration space is spanned by $\bar \phi_I\,^i$ (see \Eref{defofphi}). On the other hand, the densitized triad dual to the connection is given by 
\begin{equation}\label{defofp}
(E^a_i)=\sqrt{h}\bar p^I\,_i X^a_I=\sqrt{h}\left(
\begin{array}{ccc}
\frac{\bar p^1\,_1}{a_1\,^1} & 0 &  0 \\
0 & \frac{a_3\,^3\bar p^2\,_2-a_3\,^2\bar p^3\,_2}{\mathfrak{h}} & \frac{a_3\,^3\bar p^2\,_3-a_3\,^2\bar p^3\,_3}{\mathfrak{h}} \\
0 & \frac{a_2\,^2 \bar p^3\,_2-a_2\,^3\bar p^2\,_2}{\mathfrak{h}} & \frac{a_2\,^2\bar p^3\,_3-a_2\,^3\bar p^2\,_3}{\mathfrak{h}}
\end{array}
\right), \end{equation}
where
\ben (\bar p^I\,_i)=\left(
\begin{array}{ccc}
\bar p^1\,_1 & 0 &  0 \\
0 & \bar p^2\,_2 & \bar p^2\,_3 \\
0 & \bar p^3\,_2 & \bar p^3\,_3
\end{array}
\right),
\een
together with $\omega_a^JX^a_I=\delta^J_I$ and  $h=(a_1\,^1)^2(a_2\,^3a_3\,^2-a_2\,^2a_3\,^3)^2$ is the determinant of the spatial metric constructed from the vectors $a_i$ and $\bar p^I\,_i$ the momentum dual to $\bar\phi_I\,^i$ satisfying the Poisson bracket
\begin{equation}\label{bracketphipV0}
\{\bar\phi_I\,^i,\bar p^J\,_j\}=\frac{\kappa\gamma}{V_0}\delta_I^J\delta_j^i
\end{equation}
with the volume $V_0=\int_{\tor}d^3x\sqrt{h}$ of $\tor$ as measured by the metric $h$. For later purpose we define new variables
\begin{equation}\label{pphi}
\phi_I\,^i = L_I\bar\phi_I\,^i,\quad p^I\,_i=\frac{V_0}{L_I}\bar p^I\,_i,
\end{equation}
such that
\begin{equation}\label{bracketphip}
\{\phi_I\,^i, p^J\,_j\}=\kappa\gamma\delta_I^J\delta_j^i,
\end{equation}
where
\begin{eqnarray*}
L_1=a_1\,^1, \; L_2=\sqrt{(a_2\,^2)^2+(a_2\,^3)^2},\; L_3=\sqrt{(a_3\,^2)^2+(a_3\,^3)^2}.
\end{eqnarray*}
Thus we conclude that
\begin{Prop}
The classical configuration space $\mathcal{A}_S=\mathds{R}^{5}$ is spanned by the five configuration variables $\phi_I\,^i$. The phase space $\mathcal{P}=\mathds{R}^{10}$ is spanned by $\phi_I\,^i$ and the five momenta $p^J\,_j$ satisfying the Poisson bracket \eref{bracketphip}.
\end{Prop}
Furthermore, note that the determinant of the densitized triad is given by
\begin{equation}\label{detE}
\det E_i^a=\mathfrak{k}\, p^1\,_1( p^2\,_3 p^3\,_2- p^2\,_2 p^3\,_3),
\end{equation}
where we defined
\ben \mathfrak{k}:=\frac{L_1 L_2 L_3}{V_0}.\een

The relation between the new variables $(\phi_I\,^i,p^J\,_j)$ and the 'scale factors' $a_a\,^b$ and their respective momenta $p^a\,_b$ can be found by using \Eref{EEeqh} and the Poisson brackets \eref{bracketap} and \eref{bracketphip}. A closed form could only be found for $p^1\,_1$ and is given by
$$|p^1\,_1|=|a_2\,^2a_3\,^3-a_2\,^3a_3\,^2|.$$

\subsection{Constraints in Ashtekar Variables on the Torus}
\begin{figure}[!ht]
 \begin{center}
 \includegraphics[width=7.0cm]{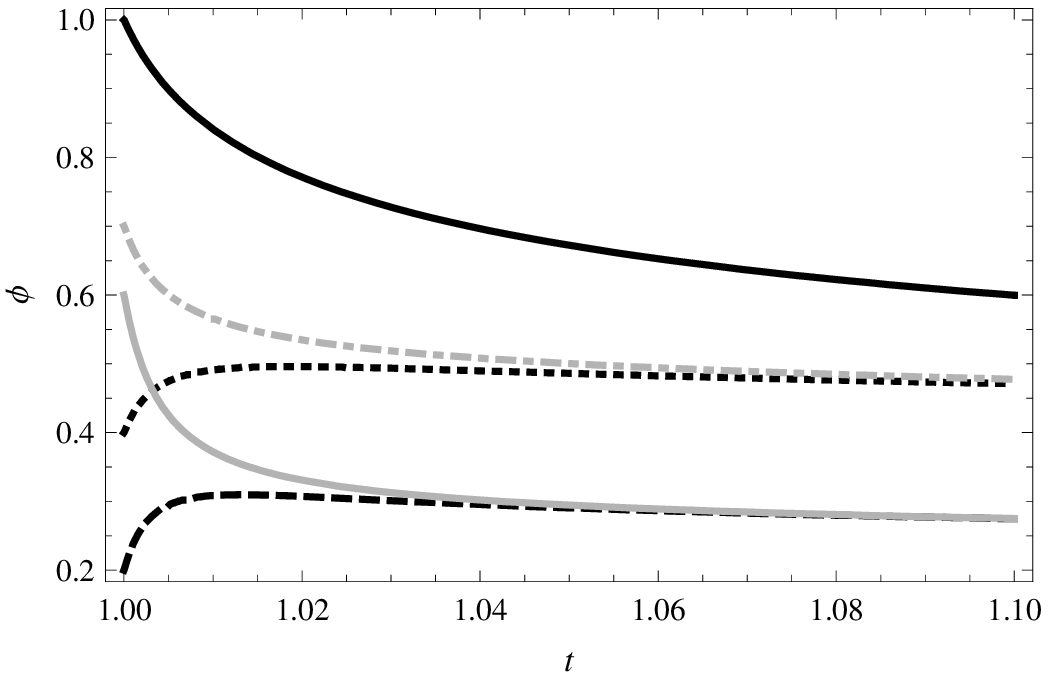}\quad\includegraphics[width=7.0cm]{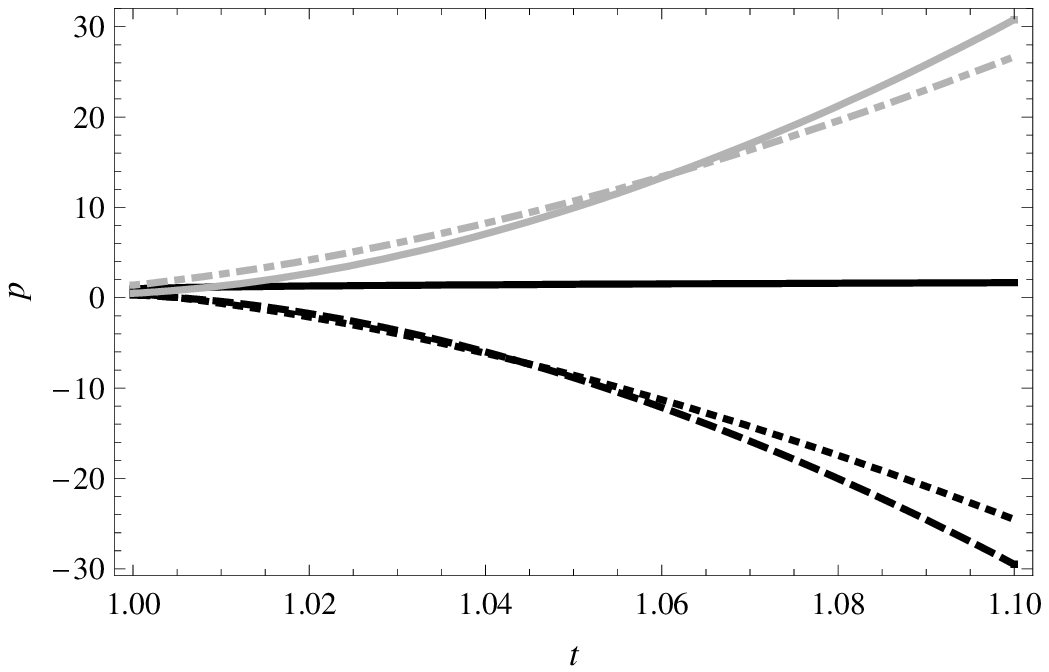}
\caption{Solutions corresponding to the Hamiltonian \eref{Cgrav} coupled to a massless scalar field with vanishing potential. {\it Left panel}: the black thick solid shows the evolution of $\phi_1\,^1$, $\phi_2\,^2$ is the black dashed line, $\phi_3\,^3$ the dotted line, $\phi_2\,^3$ the gray dashdotted one and $\phi_3\,^2$ the solid gray one. {\it Right panel}: the black thick solid shows the evolution of $p^1\,_1$, $p^2\,_2$ is the black dashed line, $p^3\,_3$ the dotted line, $p^2\,_3$ the gray dashdotted one and $p^3\,_2$ the solid gray one. In both cases the initial conditions are $\phi_1\,^1=1.0$, $\phi_2\,^2=0.2$, $\phi_3\,^3=0.4$, $\phi_2\,^3=0.6$, $\phi_3\,^2=0.7$, $p^1\,_1=1.0$, $p^2\,_2=0.3$, $p^3\,_3=0.5$, $p^2\,_30.5$, $p^3\,_2=1.4$, $\phi=0.01$ and $p_{\phi}=8.1$. The time $t$ parametrizes the coordinate time in natural units ($c=\kappa=\hbar=1$).}\label{fig:phipoft}
\end{center}
\end{figure}

In the canonical variables \eref{AEvariables} the Legendre transform of the Einstein-Hilbert action \eref{SEH} results in a fully constrained system \cite{Ashtekar:04,Rovelli:04,Thiemann:07}
\begin{equation}\label{SAE}
S=\frac{1}{2\kappa}\int_{\mathds{R}}dt\int_{\tor}d^3x\left(2\dot A_a^iE^a_i-[\Lambda^jG_j+N^aH_a+N\mathcal{H}]\right),
\end{equation}
where $G_j$ is the Gauss constraint, $H_a$ the diffeomorphism (or vector) constraint, $\mathcal{H}$ the Hamiltonian and $\Lambda^j$, $N^a$, $N$ are Lagrange multipliers. The Hamiltonian constraint simplifies to
\begin{equation}\label{Cgravdef}
C_{\mathrm{grav}}=-\frac{1}{2\kappa}\int_{\tor}d^3xN\epsilon_{ijk}F^i_{ab}\frac{E^{aj}E^{bk}}{\sqrt{|\mathrm{det}E|}}
\end{equation}
due to spatial flatness, where the curvature of the Ashtekar connection is given by
\ben F_{ab}^i=\partial_aA_b^i-\partial_bA_a^i+\epsilon^i_{jk}A^j_aA^k_b= \epsilon^i_{jk}A^j_aA^k_b. \een
Homogeneity further requires that $N\neq N(x)$. Inserting \Eref{defofphi} and \Eref{defofp} into \Eref{Cgravdef} we get
\ba \label{Cgrav}
C_{\mathrm{grav}}=&&-\frac{1}{\kappa\gamma^2}\frac{1}{\sqrt{|p^1\,_1(p^2\,_2p^3\,_3-p^2\,_3p^3\,_2)|}}\times\nonumber \\
&\Bigl[&\phi_1\,^1p^1\,_1\left\{(\phi_2\,^2-\phi_2\,^3)(p^2\,_2-p^2\,_3)+(\phi_3\,^2-\phi_3\,^3)(p^3\,_2-p^3\,_3)\right\}\nonumber\\
&&+(\phi_2\,^3\phi_3\,^2-\phi_2\,^2\phi_3\,^3)(p^2\,_3p^3\,_2-p^2\,_2p^3\,_3)\Bigr],
\ea
where we defined $N=\sqrt{L_1 L_2 L_3/V_0}$ in order to simplify the Hamiltonian. Using the Hamiltonian \eref{Cgrav} we can compute the time evolution of the basic variables $\phi_i\,^j$ and $p^i\,_j$ (see \Fref{fig:phipoft}). Setting all off-diagonal terms to zero we see that \Eref{Cgrav} matches with Eq.~(3.20) in \cite{Chiou:07}. If we further set $\phi_{(i)}\,^i=c$ and $p^{(i)}\,_i=p$ we get
$$C_{\mathrm{grav}}=-\frac{3}{\kappa\gamma^2}c^2\sqrt{|p|},$$
which is exactly the same result as the homogeneous and isotropic case \cite{Bojowald:08}.

\subsection{Diffeomeorphism and Gauss Constraints}\label{sec:diffgauss}
The Gauss constraint stems from the fact that we chose the densitized triads $E_i^a$ as the momenta conjugated to the connections $A_a^i$. In fact, the spacial metric can be directly obtained from the densitized triads through \Eref{EEeqh} and is invariant under rotations given by $E_i^a\mapsto O_i^jE_j^a$. In order that the theory be invariant under such rotations the Gauss constraint
\be\label{Gaussconstraint} G_i=\partial_aE^a_i+\epsilon_{ijk}A^j_aE^a_k\approx0 \ee
must be satisfied. The diffeomorphism constraint modulo Gauss constraint originates from the requirement of independence from any spatial coordinate system or background and is given by
\be\label{diffeoconstraint} H_a=F^i_{ab}E^b_i\approx0.\ee

However, as mentioned in \cite{Ashtekar:91} we have to be careful when dealing with these constraints in the case where the topology is closed. We thus divide this subsection into two parts, starting with the general case of open models.

\subsubsection{Open Models}\label{sec:openmodels}
Due to spatial homogeneity of Bianchi type I models the basic variables can be diagonalized to \cite{Bojowald:00,Bojowald:03}
\ben A^{'i}_a=\tilde c'_{(K)}\Lambda^{'i}_K\omega^{'K}_a,\quad E^{'a}_i=p^{'(K)}\Lambda^{'K}_iX^{'a}_K, \een
where $\omega'$ is the left-invariant 1-form, $X'$ the densitized left-invariant vector field dual to $\omega'$ and $\Lambda'\in SO(3)$\footnote{In order to avoid confusion with the rest of this work we tag every variable with a $"'"$ when dealing with the open case.}. This choice of variables automatically satisfies the vector and Gauss constraints, thus reducing the analysis of \Eref{SAE} to the Hamiltonian constraint \eref{Cgravdef}. The homogeneous, anisotropic vacuum solution to the Einstein field equations is called the Kasner solution and is given by the following metric:
\ben ds^2=-d\tau^2+\tau^{2\alpha_1}dx_1^2+\tau^{2\alpha_2}dx_2^2+\tau^{2\alpha_3}dx_3^2 \een
where the two constraints $\alpha_i\in\mathds{R}$, $\sum \alpha_i=\sum \alpha_i^2=1$ have to be fulfilled. These imply that not all Kasner exponents can be equal, i.e. isotropic expansion or contraction of space is not allowed. By contrast the RW metric is able to expand or contract isotropically because of the presence of matter. At the end, from the twelve-dimensional phase space only two degrees of freedom remains.

An infinitesimal diffeomorphism generated by a vector field $V$ induces the following action on the left-invariant 1-form $\omega'$:
\be\label{HPD} \omega_a\mapsto\omega'_a+\epsilon\mathcal{L}_V\omega'_a, \ee
where $\mathcal{L}_V$ is the Lie derivative along $V$. Such transformations leave the metric homogeneous provided the vector fields satisfy 
\be\label{condHPD} V^a=-(f_j^iy^j)X^{'a}_i \ee
for some constants $f_j^i$ and functions $y^i$ given by $\mathcal{L}_{K_j}y^i=\delta_j^i$ \cite{Ashtekar:91}. The last equation for $y^i$ relies on the fact that the 3-surface is topologically $\mathds{R}^3$ and the Killing vectors $K_i$ commute. As we shall see below this will not be the case in the closed models.

In the case of rotational symmetry the diffeomorphism constraint is once again satisfied by the choice of variables whereas the Gauss constraint is not. However, in such a case the triad components can be rotated until the Gauss constraint is also satisfied. Further details can be found in \cite{Ashtekar:06:3}.

\subsubsection{The Torus as a Closed Model}\label{sec:closedmodels}
As we have seen in \Sref{sec:CHU} it is not possible to align the Killing fields with the left-invariant vectors, whence the metric takes the non-diagonal form \eref{metrich} and the Ashtekar connection the form \eref{defofphigen}. In the previous subsection we saw that a diffeomorphism preserves homogeneity provided it satisfies the condition \eref{condHPD}. In the closed model the analysis goes through as well and we find that $V_i$ has to satisfy the same condition \eref{condHPD}. However, since such fields lack the required periodicity in $x^i$ we are led to the conclusion that there are no globally defined, non-trivial homogeneity preserving diffeomorphisms (HPDs) and there is no analog of \eref{HPD}. Thus, instead of one degree of freedom we get additional degrees of freedom.

The Gauss constraint for a Bianchi type I model is given by
\be \label{GaussBianchi} G_i=\epsilon_{ijk}\phi_I\,^ip^I\,_k.\ee 
With our choice of variables two Gauss constraints are automatically satisfied, namely $G_2=G_3\equiv0$. However, we can still perform a global $SU(2)$ transformation along $\tau_1$ which is implemented in the nonvanishing Gauss constraint
\be\label{G1} G_1=\phi_2\,^2p^2\,_3+\phi_3\,^2p^3\,_3-\phi_2\,^3p^2\,_2-\phi_3\,^3p^3\,_2\approx0 \ee
generating simultaneous rotations of the pairs $(\phi_2\,^2,\phi_2\,^3)$, $(p^2\,_2,p^2\,_3)$ resp. $(\phi_3\,^2,\phi_3\,^3)$, $(p^3\,_2,p^3\,_3)$. Thus the norms of these vectors and the scalar products between them are gauge invariant. The Gauss constraint allows us to get rid of e.g. the pair $(\phi_3\,^2,p^3\,_2)$ by fixing the gauge in the following way: we rotate the connection components such that $\phi_3\,^2=0$. Because the length $\|\phi_3\|=\sqrt{(\phi_3\,^2)^2+(\phi_3\,^3)^2}$ is preserved we know that $\phi_3\,^3\neq0$.  The Gauss constraint then implies that $p^3\,_2=(\phi_2\,^2p^2\,_3-\phi_2\,^3p^2\,_2)/\phi_3\,^3$. This gauge fixing reduces the degrees of freedom by two units.

The diffeomorphism constraint is given by \Eref{diffeoconstraint} and since $F_{ab}^i=\epsilon^i\,_{jk}A_a^jA_b^k$ ($\partial_aA_b^i=0$ thanks to homogeneity) we find that
\be\label{diffeopropGauss} H_a=\epsilon^i\,_{jk}A_a^jA_b^kE^b_i\propto A^i_aG_i. \ee
The gauge fixing we just performed ensures that the diffeomorphism constraint also vanishes.

\subsection{Canonical Transformation}\label{sec:cantransf}
In this subsection we introduce a set of new variables which will greatly simplify the analysis of the kinematical Hilbert space. We first perform a canonical transformation on the unreduced phase space:
\ba\label{cantransf}
Q_1=\phi_1\,^1,\quad&& P^1=p^1\,_1,\nonumber\\
Q_2=\sqrt{(\phi_2\,^2)^2+(\phi_2\,^3)^2},\quad&& P^2=\frac{p^2\,_2\phi_2\,^2+p^2\,_3\phi_2\,^3}{\sqrt{(\phi_2\,^2)^2+(\phi_2\,^3)^2}}\nonumber\\
Q_3=\sqrt{(\phi_3\,^2)^2+(\phi_3\,^3)^2},\quad&& P^3=\frac{p^3\,_2\phi_3\,^2+p^3\,_3\phi_3\,^3}{\sqrt{(\phi_3\,^2)^2+(\phi_3\,^3)^2}}\\
\theta_1=\mathrm{arc}_k\mathrm{cos}\left(\frac{\phi_2\,^2}{\sqrt{(\phi_2\,^2)^2+(\phi_2\,^3)^2}}\right),\quad&& P_{\theta_1}=p^2\,_3\phi_2\,^2-p^2\,_2\phi_2\,^3\nonumber\\
\theta_2=\mathrm{arc}_k\mathrm{cos}\left(\frac{\phi_3\,^3}{\sqrt{(\phi_3\,^2)^2+(\phi_3\,^3)^2}}\right),\quad&& P_{\theta_2}=-p^3\,_3\phi_3\,^2+p^3\,_2\phi_3\,^3\nonumber
\ea
such that the variables are mutually conjugate:
\ben
\{Q_I,P^J\}=\frac{\kappa\gamma}{V_0}\delta_I^J,\quad \{\theta_{\alpha},P_{\theta_{\beta}}\}=\frac{\kappa\gamma}{V_0}\delta_{\alpha,\beta}.
\een
We choose the convention that the diagonal limit can be recovered by setting $\theta_1=\theta_2=0$. The inverse of this canonical transformation will be important in the sequel and is given by:
\ba\label{invcantransf} \phi_2\,^2=Q_2\cos(\theta_1),\quad&&\phi_2\,^3= Q_2\sin(\theta_1),\nonumber\\
p^2\,_2=P^2\cos(\theta_1)-\frac{P_{\theta_1}\sin(\theta_1)}{Q_2},\quad &&p^2\,_3=\frac{P_{\theta_1}\cos(\theta_1)}{Q_2}+P^2\sin(\theta_1),\\
\phi_3\,^2= Q_3\sin(\theta_2),\quad&&\phi_3\,^3= Q_3\cos(\theta_2),\nonumber\\
p^3\,_2= P^3\sin(\theta_2)+\frac{P_{\theta_2}\cos(\theta_2)}{Q_3},\quad &&p^3\,_3=-\frac{P_{\theta_2}\sin(\theta_2)}{Q_3}+ P^3\cos(\theta_2).\nonumber
\ea
It is important to note that $Q_2,Q_3\in\mathds{R}_+$ and $\theta_1,\theta_2\in [k\pi,(k+1)\pi]$ where we restrict the values of $k$ to be either $k=0$ if $\mathrm{sgn}(\phi_2\,^3)>0$ or $k=1$ if $\mathrm{sgn}(\phi_2\,^3)<0$. If $\mathrm{sgn}(\phi_2\,^3)=0$ then we have the case $k=0$ if $\mathrm{sgn}(\phi_2\,^2)>0$ or $k=1$ if $\mathrm{sgn}(\phi_2\,^2)<0$. The function arc$_1$cos$(x)$ is related to the principal value via arc$_1$cos$(x)=2\pi-$arccos$(x)$. With this convention we can recover \Eref{cantransf} unambiguiously from \Eref{invcantransf}.

\begin{figure}[!ht]
 \begin{center}
 \includegraphics[width=7.0cm]{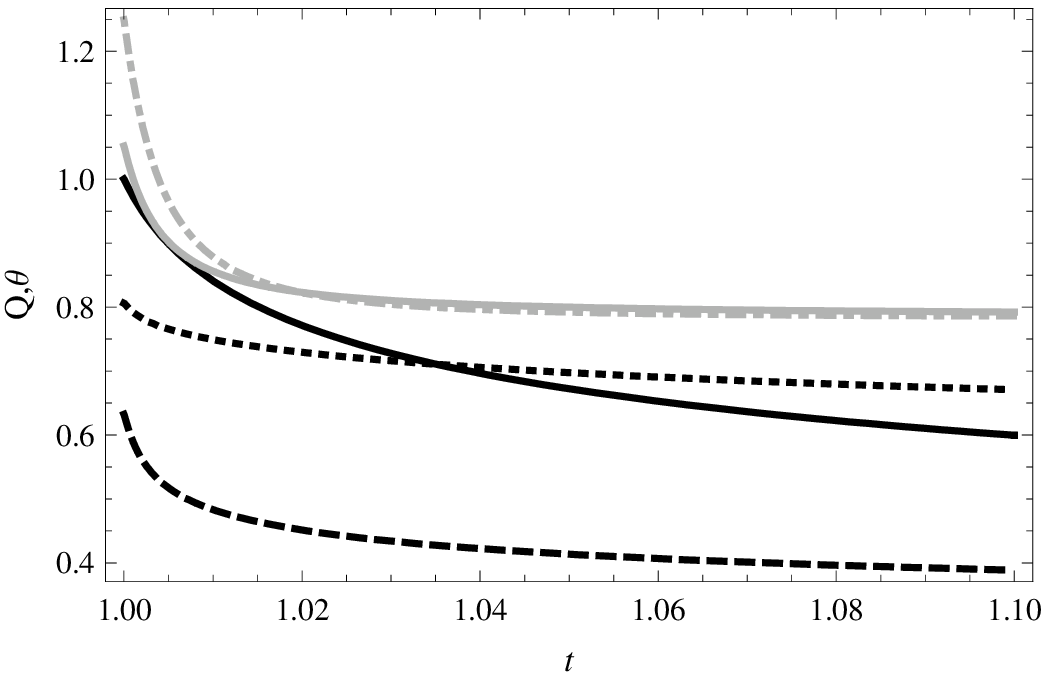}\quad\includegraphics[width=7.0cm]{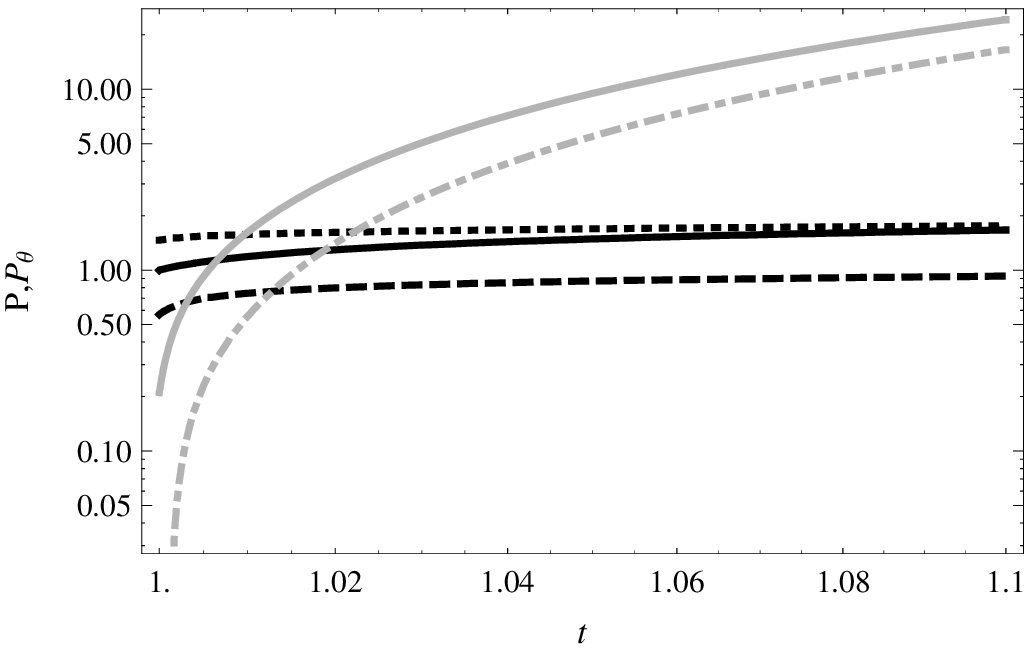}
\caption{Solutions corresponding to the Hamiltonian \eref{Cgravnew} coupled to a massless scalar field with vanishing potential. {\it Left panel}: the black thick solid shows the evolution of $Q_1$, $Q_2$ is the black dashed line, $Q_3$ the dotted line, $\theta_1$ the gray dashdotted one and $\theta_2$ the solid gray one. {\it Right panel}: the black thick solid shows the evolution of $P^1$, $P^2$ is the black dashed line, $P^3$ the dotted line, $P_{\theta_1}$ the gray dashdotted one and $P_{\theta_2}$ the solid gray one. In both cases the initial conditions are $Q_1=1$, $Q_2=0.63$, $Q_3=0.81$, $\theta_1=1.25$, $\theta_2=1.05$, $P^1=1$, $P^2=0.57$, $P^3=1.46$, $P_{\theta_1}=-0.08$, $P_{\theta_2}=0.21$, $\phi=0.01$ and $p_{\phi}=8.1$. The time $t$ parametrizes the coordinate time in natural units ($c=\kappa=\hbar=1$).}\label{fig:QPoft}
\end{center}
\end{figure}

The Hamiltonian constraint \eref{Cgrav} is given in terms of the new variables by
\ba\label{Cgravnew}
C_{\mathrm{grav}}=\frac{(2\kappa\gamma^2)^{-1}}{\sqrt{\left|\frac{P^1\left[\cos(\theta_1+\theta_2)(P_{\theta_1}P_{\theta_2}-P^2P^3Q_2Q_3)+(P^2P_{\theta_2}Q_2+P^3P_{\theta_1}Q_3)\sin(\theta_1+\theta_2)\right]}{Q_2Q_3}\right|}}\times\nonumber\\
\times\biggl\{2P^1Q_1\Bigl[\cos(2\theta_2)P_{\theta_2}+P^2Q_2(\sin(2\theta_1)-1)+P^3Q_3(\sin(2\theta_2)-1)\Bigr]\nonumber\\
\quad\;\;+P^2Q_2\Bigl[P_{\theta_2}\sin(2(\theta_1+\theta_2))-2\cos^2(\theta_1+\theta_2)P^3Q_3\Bigr]\\
\quad\;\; +P_{\theta_1}\Bigl[2\cos^2(\theta_1+\theta_2)P_{\theta_2}+2\cos(2\theta_1)P^1Q_1+P^3P_{\theta_3}\sin(2(\theta_1+\theta_2))\Bigr]\biggl\}\nonumber
\ea
Using this Hamiltonian we can compute the time evolution of the basic variables $Q_i$, $\theta_{\alpha}$, $P^i$ and $P_{\theta_{\alpha}}$ (see \Fref{fig:QPoft}). We choose the initial conditions so that they correspond to the values of the old variables (see caption of \Fref{fig:phipoft}). By doing so we are able to check whether the solutions to \Eref{Cgrav} are equivalent to the solutions to \Eref{Cgravnew} by performing the canonical transformation \eref{cantransf}. The different solutions do indeed match up to a very good accuracy.

 The only nontrivial Gauss constraint \eref{G1} is then given by
\be\label{G1inPQ}
G_1=P_{\theta_1}-P_{\theta_2},
\ee
which vanishes only when $P_{\theta_2}=P_{\theta_1}$. We are free to fix the gauge by setting $\theta_2=0$. The same result can be obtained from the gauge fixing performed in \Sref{sec:closedmodels} so that
$$Q_3=\phi_3\,^3,\quad P^3=p^3\,_3,\quad \theta_2=0\quad\mathrm{and}\quad P_{\theta_2}=P_{\theta_1}.$$
The symplectic structure of the reduced 8-dimensional phase space is given by
$$\mathbf{\Omega}=\frac{V_0}{\kappa\gamma}(dQ_1\wedge dP^1+dQ_2\wedge dP^2+dQ_3\wedge dP^3+d\theta_1\wedge dP_{\theta_1}).$$

\section{Kinematical Hilbert Space}\label{sec:Hkin}
\subsection{Holonomies}\label{sec:holonomyfluxalgebra}
In the last section we parametrized the classical phase space and gave the Hamiltonian in terms of the Ashtekar variables. To quantize the theory we have to select a set of elementary observables which have unambiguous operator analogs. In order to do so we have first to find elementary variables of the 5-dimensional configuration space.

According to \cite{Bojowald:00:2} the configuration space on the covering space $\tilde \Sigma$ is given by Higgs fields in a single point $x_0$ which is the only point in the reduced manifold $\tilde \Sigma/\tilde S$. In quantum theory these fields are represented as point holonomies associated to the point $x_0$ \cite{Thiemann:98:4}. On $\tilde S$ we can take the three edges $\xi_1$, $\xi_2$ and $\xi_3$ in order to regularize the point holonomies. However, we would like to apply this construction to a closed manifold. First we note that $\Sigma\cong \tilde S/\Gamma$ such that two elements $g,g'\in \tilde S$ are equivalent if there is an element $\gamma\in\Gamma$ such that $g'=g+\gamma$. We can thus restrict the regularization of the point holomonies to the three edges $X_1$, $X_2$ and $X_3$ meeting at $x_0$ without losing information.  Our elementary configuration variables are then the holonomies along straight lines $\gamma_I:[0,1]\rightarrow \Sigma$ defined by the connection $A \propto (\phi_I\,^i)$ \cite{Bojowald:00:2,Bojowald:00:3,Bojowald:02,Ashtekar:03,Bojowald:03}. Now the holonomies along $X_1$, $X_2$ resp. $X_3$ are given by
\begin{eqnarray}\label{holonomiesalonga}
h_1^{(\lambda_1)}&=&\exp(\lambda_1\phi_1\,^1\tau_1)=\cos(\lambda_1\phi_1\,^1)+2\tau_1\sin(\lambda_1\phi_1\,^1)\nonumber\\
h_2^{(\lambda_2)}&=&\exp(\lambda_2\phi_2\,^2\tau_2+\lambda_2\phi_2\,^3\tau_3)\\
h_3^{(\lambda_3)}&=&\exp(\lambda_3\phi_3\,^2\tau_2+\lambda_3\phi_3\,^3\tau_3), \nonumber
\end{eqnarray}
where $\lambda_I\in(-\infty,\infty)$ and $\lambda_IL_I$ is the length of the edge $I$ with respect to the spatial metric $h_{ab}$. The auxilary Hilbert space is then generated by spin networks associated with graphs consisting of the three edges $\gamma_I$ meeting at the vertex $x_0$.

In open Bianchi type I models the gauge invariant information of the connection can be separated from the gauge degrees of freedom via the relation $\phi_I^i=c_{(I)}\Lambda_I^i$ with $\Lambda\in SO(3)$ so that the holonomies become simple trigonometric functions. In our case the situation is more complicated because the holonomies $h_2$ and $h_3$ cannot be reduced to such functions since
\ben
h_{\alpha}^{(\lambda_{\alpha})}=\cos\left(\lambda_{\alpha}\|\vec\phi_{\alpha}\|/2\right)+2\frac{\phi_{\alpha}\,^i\tau_i}{\|\vec\phi_{\alpha}\|}\sin\left(\lambda_{\alpha}\|\vec\phi_{\alpha}\|/2\right)\quad(\mathrm{no}\;\mathrm{summation}), \een
where $\alpha=2,3$ and 
\ben
\|\vec\phi_{\alpha}\|:=\sqrt{\sum_{i}(\phi_{\alpha}\,^i)^2}.\een
The problem is that this expression cannot be used in this form since there is no well defined operator $\hat \phi_I\,^i$ on the kinematical Hilbert space. Using the canonical transformation \eref{cantransf} we can re-express the holonomies such that
\ba\label{holinQP}
h_1^{(\lambda_1)}=\cos(\lambda_1Q_1/2)+2\tau_1\sin(\lambda_1Q_1/2),\nonumber\\
h_2^{(\lambda_2)}=\cos(\lambda_2Q_2/2)+2(\tau_2\cos \theta_1+\tau_3\sin \theta_1)\sin(\lambda_2Q_2/2),\\
h_3^{(\lambda_3)}=\cos(\lambda_3Q_3/2)+2(\tau_2\sin \theta_2+\tau_3\cos \theta_2)\sin(\lambda_3Q_3/2).\nonumber\ea
Since $\lambda_I\in\mathds{R}$ matrix elements of the exponentials of $Q_1$, $Q_2$ and $Q_3$ form a $C^*$-algebra of almost periodic functions. On the other hand the variables $\theta_{1,2}$ are periodic angles such that only strictly periodic functions $\exp(ik_{\alpha}\theta_{\alpha})\in U(1)$ with $k_{\alpha}\in\mathds{Z}$ are allowed. Thus, any function generated by this set can be written as
\ba\label{gofQP} g(Q_1,Q_2,Q_3,\theta_1,\theta_2)=\sum_{\lambda_1,\lambda_2,\lambda_3,k_1,k_2}\xi_{\lambda_1,\lambda_2,\lambda_3,k_1,k_2}\times \nonumber\\ \quad\quad\times\exp\left(\frac{1}{2}i\lambda_1Q_1+\frac{1}{2}i\lambda_2Q_2+\frac{1}{2}i\lambda_3Q_3+ik_1\theta_1+ik_2\theta_2\right)\ea
with coefficients $\xi_{\lambda_1,\lambda_2,\lambda_3,k_1,k_2}\in\mathds{C}$, generating the $C^*$-algebra $\mathcal{A}_S$. Note that this function is almost periodic in $Q_1$,$ Q_2$ and $Q_3$ and strictly periodic in $\theta_1$ and $\theta_2$. The spectrum of the algebra of the almost periodic functions is called the Bohr compactification $\bar \mathds{R}_B:=\Delta(\Cyl)$ of the real line and can be seen as the space of generalized connections \cite{Ashtekar:03,Velhinho:07}. Thus the functions \eref{gofQP} provide us a complete set of continuous functions on $\bar\mathds{R}_B\times\bar\mathds{R}_B\times\bar\mathds{R}_B\times S^1\times S^1$. Moreover the Gel'fand theory guarantees that the space $\bar\mathds{R}_B$ is compact and Hausdorff \cite{Bratteli:79} with a unique normalized Haar measure $d\mu(c)$ such that
\ben \int f(c)d\mu(c):=\lim_{T\rightarrow\infty}\frac{1}{2T}\int_{-T}^Tf(c)dc.\een

A Cauchy completion leads to a Hilbert space $\HS$ defined by the tensor product $\HS=\mathcal{H}_B^{\otimes3}\otimes\mathcal{H}_{S^1}^{\otimes2}$ with the Hilbert spaces $\mathcal{H}_B=L^2(\bar\mathds{R}_B,d\mu(c))$ and $\mathcal{H}_{S^1}=L^2(S^1,d\phi)$ of square integrable functions on $\bar\mathds{R}_B$ and the circle respectively, where $d\phi$ is the Haar measure for $S^1$. An orthonormal basis for $\mathcal{H}_B$ is given by the almost periodic functions $\langle Q_I|\mu_I\rangle=\exp(i\mu_IQ_I/2)$ (no summation) with $\mu_I\in\mathds{R}$ with $\langle\mu_I|\mu_I'\rangle=\delta_{\mu_I,\mu_I'}$. Analogously a basis for $\mathcal{H}_{S^1}$ is given by the strictly periodic functions $\langle\theta_{\alpha}|k_{\alpha}\rangle=\exp(ik_{\alpha}\theta_{\alpha})$ with $\langle k_{\alpha}|k'_{\alpha}\rangle=\delta_{k_{\alpha},k'_{\alpha}}$.

We choose a representation where the configuration variables, now promoted to operators, act by multiplication via:
\ben(\hat g_1g_2)(\vec Q,\vec\theta)=g_1(\vec Q,\vec\theta)g_2(\vec Q,\vec\theta).\een
The momentum operators act by derivation in the following way:
\be\label{Pop} \hat P^I=-i\gamma\lpl^2\frac{\partial}{\partial Q_I},\quad \hat P_{\theta_{\alpha}}=-i\gamma\lpl^2\frac{\partial}{\partial \theta_{\alpha}}.\ee
The eigenstates of all momentum operators are given by
\ban |\vec\mu,\vec k\rangle&:=&|\mu_1,\mu_2,\mu_3,k_1,k_2\rangle\nonumber\\
&:=&|\mu_1\rangle\otimes|\mu_2\rangle\otimes|\mu_3\rangle\otimes|k_1\rangle\otimes|k_2\rangle\ean
with
\be\label{eigenvaluesofP} \hat P^I|\vec\mu,\vec k\rangle=\gamma \lpl^2\mu_I|\vec\mu,\vec k\rangle,\quad \hat P_{\theta_{\alpha}}|\vec\mu,\vec k\rangle=\frac{\gamma \lpl^2}{2}k_{\alpha}|\vec\mu,\vec k\rangle.\ee

The simple form of the momentum operators \eref{Pop} may suggest that the Hilbert space of LQC on a torus is simply expanded from $L^2(\bar\mathds{R}_B^3)$ to $L^2(\bar\mathds{R}_B^3)\times L^2(U(1)^2)$. However the situation is far more complicated because the important variables for the Gauss and Hamiltonian constraints are not the new momenta $P^I$ and $P_{\theta_{\alpha}}$ but the components $p^I\,_i$ of the triad. In terms of the new canonical variables they are complicated functions of both the configuration and momentum variables, as can be seen from \Eref{invcantransf}. These expressions cannot be quantized directly since the operators $\hat Q_{2,3}$ fail to be well defined on the Hilbert space. The solution is to consider the momentum operators of the full theory given by a sum of left and right invariant vector fields. In \cite{Bojowald:00:3} the same strategy was used to show that the triad components $p^I\,_i$ act by derivation. In our case the situation is more complicated since the triad components contain both configuration and momentum variables. The triad operators act on functions in $\HS$ and are given by
\be\label{defofpX} \hat p^I\,_i=-i\frac{\gamma\lpl^2}{2}\left(X^{(R)}_i(h_I)+X^{(L)}_i(h_I)\right),\ee
where $X^{(R)}_i(h_I)$ and $X^{(L)}_i(h_I)$ are the right and left invariant vector fields acting on the copy of $SU(2)$ associated with the edge $e_I$ of length 1 and are given by
\ben X^{(R)}_i(h_I)=\mathrm{tr}\left[(\tau_ih_I)^T\frac{\partial}{\partial h_I}\right],\quad X^{(L)}_i(h_I)=\mathrm{tr}\left[(h_I\tau_i)^T\frac{\partial}{\partial h_I}\right].\een
Applying the operators $\hat p^2\,_2$ and $\hat p^2\,_3$ on the function $\tr(h_2)$ we get
\ba\label{operatorp} \hat p^2\,_2\tr(h_2)=2\hat p^2\,_2\cos(\lambda_2Q_2/2)=i\gamma\lpl^2\lambda_2\sin(\lambda_2Q_2/2)\cos(\theta_1),\nonumber\\
\hat p^2\,_3\tr(h_2)=2\hat p^2\,_3\cos(\lambda_2Q_2/2)=i\gamma\lpl^2\lambda_2\sin(\lambda_2Q_2/2)\sin(\theta_1).
\ea
We see that the usual expressions for an open topology can be recovered by simply setting $\theta_1=0$. Applying these operators once again we get the expressions:
\ben (\hat p^2\,_2)^2\tr(h_2)=\frac{1}{2}\gamma^2\lpl^4\lambda_2^2\cos(\lambda_2Q_2/2)=(\hat p^2\,_3)^2\tr(h_2), \een
which means that $\cos(\lambda_2Q_2/2)$ is an eigenfunction of both $(\hat p^2\,_2)^2$ and $(\hat p^2\,_3)^2$ with eigenvalue $\gamma^2\lpl^4/2\lambda_2^2$. On the other hand we have
\ben \hat p^2\,_2 \hat p^2\,_3\tr(h_2)=\hat p^2\,_3 \hat p^2\,_2\tr(h_2)=0.\een

\subsection{Quantization: 1. Possibility}\label{sec:loopquantization}
As previously mentioned we cannot directly quantize the expressions \eref{invcantransf} because $\hat Q_I$ does not exist as multiplication operator on $\HS$. In a loop quantization only holonomies of the connections are represented as well-defined operators on $\HS$. Thus we replace every configuration variable $Q_I$ in \Eref{invcantransf} by $\sin(\delta_I Q_I/2)/\delta_I$ \cite{Bojowald:04}, where $\delta_I\in\mathds{R}\backslash\{0\}$ plays the role of a regulator, and compare it with the results just obtained in terms of left and right invariant vector fields. For later purpose we order the operators in a symmetrical way get the following operators acting on functions of $\HS$:
\ba\label{operators} \hat\phi_2\,^2=\frac{\sin(\delta_2 Q_2)}{\delta_2}\cos\theta_1,\quad&&\hat p^2\,_2=\cos\theta_1\hat P^2-\frac{\delta_2\sqrt{\sin\theta_1}}{\sin(\delta_2 Q_2)}\hat P_{\theta_1}\sqrt{\sin\theta_1},\nonumber\\
\hat\phi_2\,^3= \frac{\sin(\delta_2 Q_2)}{\delta_2}\sin\theta_1,\quad &&\hat p^2\,_3=\sin\theta_1\hat P^2+\frac{\delta_2\sqrt{\cos\theta_1}}{\sin(\delta_2 Q_2)}\hat P_{\theta_1}\sqrt{\cos\theta_1},\nonumber\\
\hat\phi_3\,^2= \frac{\sin(\delta_3 Q_3)}{\delta_3}\sin\theta_2,\quad&&\hat p^3\,_2= \sin\theta_2\hat P^3+\frac{\delta_3\sqrt{\cos\theta_2}}{\sin(\delta_3 Q_3)}\hat P_{\theta_2}\sqrt{\cos\theta_2},\\
\hat\phi_3\,^3= \frac{\sin(\delta_3 Q_3)}{\delta_3}\cos\theta_2,\quad &&\hat p^3\,_3= \cos\theta_2\hat P^3-\frac{\delta_3\sqrt{\sin\theta_2}}{\sin(\delta_3 Q_3)}\hat P_{\theta_2}\sqrt{\sin\theta_2}.\nonumber\ea
Applying e.g. the operator $\hat p^2\,_2$ on $\cos(\lambda_2Q_2/2)$ with the definitions \eref{eigenvaluesofP} we see that we obtain the same result as \Eref{operatorp} for $\delta=1$. This is not surprising in view of the fact that we defined the operator $\hat p^I\,_i$ in \Eref{defofpX} with holonomies along edges $e_I$ of length 1.

This substitution is problematic since the configuration variables $Q_{2,3}$ are by definition positive (see \Eref{cantransf}). Therefore, for $Q_{2,3}\rightarrow \sin(\delta_{2,3}Q_{2,3})/\delta_{2,3}$ to be valid we restrict the analysis to the domain $0<Q_{2,3}<\pi$. In the diagonal case the situation is less problematic because the configuration variable $c$ is arbitrary such that $\sin(\delta c)$ is also allowed to be negative.

Classically, since the change of variables \eref{invcantransf} is a canonical transformation the symplectic structure is conserved, i.e. the Poisson bracket between $p^2\,_2$ and $p^2\,_3$ vanishes:
$$\left\{p^2\,_2,p^2\,_3\right\}_{Q,P}=0$$
A quantization of the above expression is obtained with the substitution $\{,\}\rightarrow -i[,]\hbar$ such that the commutator between $\hat p^2\,_2$ and $\hat p^2\,_3$ should also vanish. However, the consequence of the substitution of $1/Q_I$ by $\delta_I/\sin(\delta_I Q_I)$ is that commutator between these two variables doesn't vanish anymore:
\be\label{commp22p23}\left[\hat p^2\,_2,\hat p^2\,_3\right]f(Q_2,\theta_1)=-\gamma^2\lpl^4\delta_2\frac{\cos(\delta_2Q_2)-\delta_2}{\sin^2(\delta_2Q_2)}\frac{\partial f}{\partial\theta_1}\ee
Formally we can recover the classical limit by taking the limit $$\lim_{\delta_2\rightarrow0}[\hat p^2\,_2,\hat p^2\,_3]f(Q_2,\theta_1)=0,$$ which however fails to exist on $\HS$.

The operators $\hat p^I\,_i$ are partial differential operators with periodic coefficients in both $\theta$ and $Q$. In spherically symmetric quantum geometry a similar situation arises when considering the quantization of a nondiagonal triad component \cite{Bojowald:04}. However the expression of this component reduces to a Hamiltonian whose eigenvalues are discrete. In our case the situation is more complicated.

\subsubsection{Quantization of $p^2\,_2$}

In order to find eigenfunctions of the triad operators let us consider an operator of the form
\ben \hat A_{\delta}:=-i\cos\theta\frac{\partial}{\partial Q}+i\frac{\delta\sqrt{\sin(\theta)}}{\sin(\delta Q)}\frac{\partial}{\partial\theta}\sqrt{\sin\theta}.\een
A substitution $\xi=\delta Q$ shows that $\hat A_{\delta}=\delta \hat A_1\equiv\delta \hat A$ so that it is sufficient to determine the spectrum for $\delta=1$. This operator is symmetric on $\mathcal{H}_A:=L^2(\bar{\mathds{R}}_B,d\mu_B)\otimes L^2(U(1))$:
$$\langle f,\hat A g\rangle=\langle\hat Af,g\rangle,\quad\forall f,g\in\mathcal{D}(\hat A),$$
where $\mathcal{D}(\hat A)\subset\mathcal{H}_A$ is the domain of $\hat A$. The eigenfunctions of $\hat A$ are obtained by solving $\hat A f_{\lambda}(\xi,\theta)=\lambda f_{\lambda}(\xi,\theta)$, i.e.
\be\label{PDE} -i\cos\theta\frac{\partial f_{\lambda}(\xi,\theta)}{\partial\xi}+i\frac{\sin\theta}{\sin\xi}\frac{\partial f_{\lambda}(\xi,\theta)}{\partial\theta}+\frac{i}{2}\frac{\cos\theta}{\sin\xi}f_{\lambda}(\xi,\theta)=\lambda f_{\lambda}(\xi,\theta),\ee
where we constrain $\xi$ to be in the interval $[0,\pi]$ in order to avoid negative values of $\sin\xi$. We look for a solution of the form $w=w(\xi,\theta)$ \cite{Kamke:79} satisfying
\ben -i\cos\theta\frac{\partial w}{\partial\xi}+i\frac{\sin\theta}{\sin\xi}\frac{\partial w}{\partial\theta}=\left(\lambda -\frac{i}{2}\frac{\cos\theta}{\sin\xi}\right)f_{\lambda}\frac{\partial w}{\partial f_{\lambda}}\een
such that the characteristic functions are given by
\be\label{chareq} \dot{\xi}=-i\cos\theta(t),\quad\dot{\theta}=i\frac{\sin\theta(t)}{\sin\xi(t)}\quad\mathrm{and}\quad \dot{f}_{\lambda}=\left(\lambda -\frac{i}{2}\frac{\cos\theta(t)}{\sin\xi(t)}\right) f_{\lambda}(t),\ee
where the dot is the time derivative. Combining the first two equations gives after integration
\be\label{C1} \sin\theta\tan\frac{\xi}{2}=C_1,\ee
meaning that every $C^1$-function $\Omega_1(\sin\theta\tan(\xi/2))$ solves the left-hand side of \Eref{PDE}. In order to solve \Eref{PDE} for $\lambda\neq0$ we first note that 
\be\label{cosoft}\cos\theta(t)=\cos(\arcsin(C_1\cot(\xi/2)))=\sqrt{1-C_1^2\cot^2(\xi/2)}\equiv i\dot{\xi}.\ee
An integration of this equation gives the result
\be\label{solfort} t=-i\frac{\sqrt{2}b\log(\sqrt{2}a\cos(\xi/2)+b)}{a\sqrt{1-C_1^2\cot^2(\xi/2)}|\sin(\xi/2)|},\ee
where
$$a=\sqrt{1+C_1^2} \quad\mathrm{and}\quad b=\sqrt{-1+C_1^2+\cos\xi(1+C_1^2)}.$$
The last characteristic equation in \eref{chareq} can be written as
$$\dot f_{\lambda}=\frac{\partial f_{\lambda}}{\partial\xi}\dot\xi=\left(\lambda-\frac{i}{2}\frac{\cos\theta}{\sin\xi}\right)f_{\lambda}$$
such that
$$\frac{\partial f_{\lambda}}{\partial\xi}=\left(i\frac{\lambda}{\cos\theta}+\frac{1}{2\sin\xi}\right)f_{\lambda}.$$
\Eref{cosoft} can be inserted into the last equation such that after an integration we get the result
$$\log f_{\lambda}=\lambda t+\log\left(\sqrt{\tan(\xi/2)}\right)+C,$$
where $t$ is given by \Eref{solfort} and $C$ is an integration constant. The final solution to the  PDE \eref{PDE} is thus given by
\ba\label{flambda} f_{\lambda}(\xi,\theta)=&&\mathcal{N}_1\sqrt{\tan(\xi/2)}\times\nonumber\\
&&\times\left(\sqrt{2}\cos(\xi/2)\alpha_1+\beta_1\right)^{-i\frac{\sqrt{2}\lambda \beta_1}{\alpha_1|\sin(\xi/2)|\cos\theta}}\Omega_1(\sin\theta\tan(\xi/2)),\ea
where
\ban \alpha_1(\xi,\theta)&=&\sqrt{1+\sin^2\theta\tan^2(\xi/2)}\quad\mathrm{and}\nonumber\\
\beta_1(\xi,\theta)&=&\sqrt{-1+\cos\xi+(1+\cos\xi)\tan^2(\xi/2)\sin^2\theta}.
\ean
The $C^1$-function $\Omega_1(\sin\theta\tan(\xi/2))$ can be determined by e.g. boundary conditions. For simplicity we set $\Omega_1(\sin\theta\tan(\xi/2))\equiv1$ subsequently. As a cross-check we see that the first line of \Eref{flambda} solves
$$-i\cos\theta\frac{\partial\sqrt{\tan(\xi/2)}}{\partial\xi}+\frac{i}{2}\frac{\cos\theta}{\sin\xi}\sqrt{\tan(\xi/2)}=0$$
and the second one the eigenvalue problem of the operator $\hat A$. The scalar product on $\mathcal{H}_A$ is given by
\be \label{scalarproductH}\langle f_{\lambda},f_{\lambda'}\rangle=\lim_{T\rightarrow\infty}\frac{1}{2T}\int_{-T}^{T}d\xi\int_0^{2\pi}d\theta\bar f_{\lambda}f_{\lambda'}.\ee
The integral of $|\sqrt{\tan(\xi/2)}|^2$ over one period is not finite and since the second line of \Eref{flambda} never vanishes the eigenfunctions $f_{\lambda}$ are not normalizable in $\mathcal{H}_A$. We could choose the function $\Omega_1\propto(\tan(\xi/2))^{-1/2}$ but we would automatically get the factor $(\sin\theta)^{-1/2}$ which is also not normalizable. The surprising implication is that the spectrum of $\hat A$ is continuous. Note that the function $\alpha_1$ is always real while $\beta_1$ is always purely imaginary ($\lim_{\xi\rightarrow\pi/2}\beta_1=i\cos\theta)$. The exponent of $f_{\lambda}$ is thus always real, implying that $f_{\lambda}$ is uniquely determined.

\subsubsection{Self-adjointness of $\hat A$}\label{SAofA}
In the previous section we constructed a symmetric operator $\hat A$ with respect to the scalar product of $\mathcal{H}_A$, i.e. $\hat A=\hat A^+$ with domain $\mathcal{D}(\hat A)\subset\mathcal{D}(\hat A^+)$. In this subsection we give a possible domain for $\hat A$ and check if there exists a self-adjoint extension of $\hat A$.
\begin{Def}
In analogy with \cite{Shubin:74,Shubin:78} define the space CAP$(\mathds{R})$ of the (uniform) almost periodic functions\footnote{An almost periodic function $f(x)$ is uniformly continuous for $x\in\mathds{R}$ and bounded \cite{Bohr:47}.} such that its completion is the Hilbert space $L^2(\bar{\mathds{R}}_B)$. The Sobolev space $H^1(\bar{\mathds{R}}_B)$ is given by the completion of the space of trigonometric polynomials Trig$(\mathds{R})$ in the Sobolev norm $\|f\|_{H^1}^2=\|f\|^2_{L^2(\bar{\mathds{R}}_B)}+\|f'\|^2_{L^2(\bar{\mathds{R}}_B)}$, i.e. $H^1(\bar{\mathds{R}}_B)$ consists of all almost periodic functions $f\in \mathrm{CAP}(\mathds{R})$ such that $f'\in \mathrm{CAP}(\mathds{R})$.
\end{Def}
Let the differential operator $\hat p:=-i\frac{d}{d\xi}$ on $L^2(\bar{\mathds{R}}_B)$ have the domain of definition Trig$(\mathds{R})$. Then its closure has the domain $H^1(\bar{\mathds{R}}_B)$. The adjoint operator to $\hat p$ on $L^2(\bar{\mathds{R}}_B)$ has also the domain $H^1(\bar{\mathds{R}}_B)$ and coincides with $\hat p^+$ on it. Since $\hat p=\hat p^+$, $\hat p$ is essentially self-adjoint on Trig$(\mathds{R})$ \cite{Shubin:74,Shubin:78}.

Since every almost periodic function $f(x)$ is bounded a necessary condition for the inverse $f^{-1}(x)$ to be almost periodic is that $\min_x |f(x)|\neq0$. It follows that $\sin^{-1}\xi$ is not an almost periodic function. We thus define the domain
\be \label{domainofA}\mathcal{D}(\hat A):=\{\varphi\in H^1(\bar{\mathds{R}}_B)\otimes H^1(U(1))|\varphi(k\pi,\theta)=0=\varphi'(k\pi,\theta),\,k\in\mathds{Z}\},\ee
which, according to \cite{Roberts:66,Roberts:66:2}, is dense. Any function $\varphi\in\mathcal{D}(\hat A)$ removes the pole caused by $\sin^{-1}\xi$, i.e. we require that $\lim_{\xi\rightarrow k\pi} \varphi(\xi)(\sin\xi)^{-1}=0$ and $\lim_{\xi\rightarrow k\pi} \varphi'(\xi)(\sin\xi)^{-1}=0$. $k\in\mathds{Z}$. On the other hand, thanks to $\sin\theta$ in front of the differential operator $i\partial/\partial\theta$, the boundary term of an integration by part is automatically annihilated so that no boundary conditions on $\theta$ have to be imposed. Moreover the deficiency indices $n_{\pm}$ for $\hat A$ are defined by
$$n_{\pm}:=\mathrm{dim}\,\mathrm{ker}(\hat A^+\mp i).$$
The solutions to this equation do not lie in $\mathcal{D}(\hat A^+)$ such that $n_{\pm}=0$. It follows that the operator $\hat A$ is essentially self-adjoint.

\subsubsection{Quantization of $p^2\,_3$}
The eigenfunctions of $\hat p^2\,_3$ can be obtained by applying the same procedure on the symmetrized operator
$$\hat B:=-i\sin\theta\frac{\partial}{\partial \xi}-i\frac{\sqrt{\cos\theta}}{\sin\xi}\frac{\partial}{\partial\theta}\sqrt{\cos\theta}$$
The eigenfunctions $g_{\lambda}(\xi,\theta)$ are given by
\ba\label{glambda} g_{\lambda}(\xi,\theta)=\\ \quad\frac{\mathcal{N}_2}{\sqrt{\tan(\xi/2)}}\left(\sqrt{2}\cos(\xi/2)\alpha_2+\beta_2\right)^{-i\frac{\sqrt{2}\lambda \beta_2}{\alpha_2|\sin(\xi/2)|\sin\theta}}\Omega_2(\cos\theta\tan(\xi/2))\nonumber,\ea
where
\ban \alpha_2&=&\sqrt{1-\frac{\cot^2(\xi/2)}{\cos^2\theta}}\quad\mathrm{and}\nonumber\\
\beta_2&=&\sqrt{-1+\cos\xi+\frac{\cot^2(\xi/2)}{\cos^2\theta}(\cos\xi-1)}
\ean
and $\Omega_2$ is any $C^1$-function that can be determined by boundary conditions. While the function $\beta_2$ is always purely imaginary the function $\alpha_2$ is only real when $\cot^2\xi/2<\cos^2\theta$. This means that $g_{\lambda}$ is not uniquely determined. We can write $g_{\lambda}$ as 
$$g_{\lambda}(\xi,\theta)=\frac{k_2}{\sqrt{\tan(\xi/2)}}e^{F_1(\xi,\theta)\ln F_2(\xi,\theta)}$$
with the logarithm is defined by $\ln F_2=\mathrm{Ln} F_2+2\pi in$, where $n\in \mathds{Z}$ and $\mathrm{Ln}$ is the principal value of the logarithm. Inserting this solution into the eigenvalue problem $\hat B g_{\lambda}=\lambda g_{\lambda}$ it can be shown that there is only a solution for $n=0$. The eigenfunctions $g_{\lambda}$ are not normalizable since the integral of $1/|\tan(\xi/2)|$ over one period is not finite. 

As for $f_{\lambda}$ we are led to the conclusion that the spectrum of $\hat B$ is continuous. We can construct a dense subspace along the lines described in \Sref{SAofA}, the only difference being that $g_{\lambda}$ has poles at $\xi=2k\pi$ and $\theta=(2k+1)\pi/2$ whereas $f_{\lambda}$ has poles at $\xi=(2k+1)\pi$, $k\in\mathds{Z}$.

\begin{figure}[!ht]
 \begin{center}
 \includegraphics[width=7.2cm]{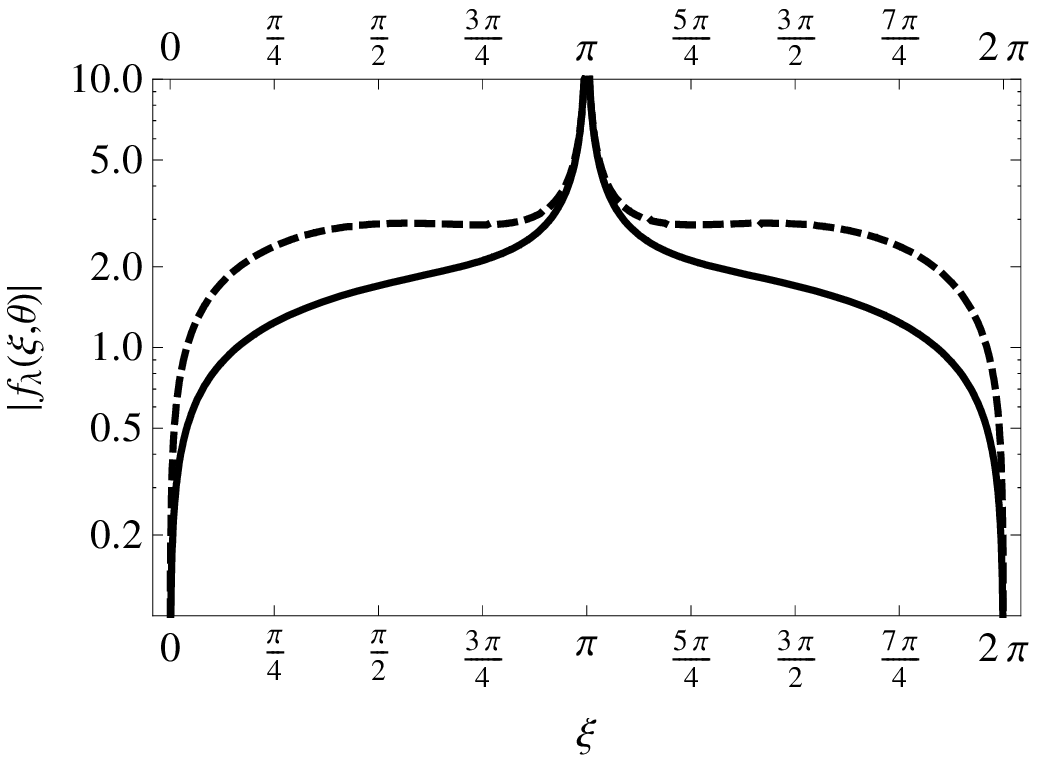}\quad\includegraphics[width=7.2cm]{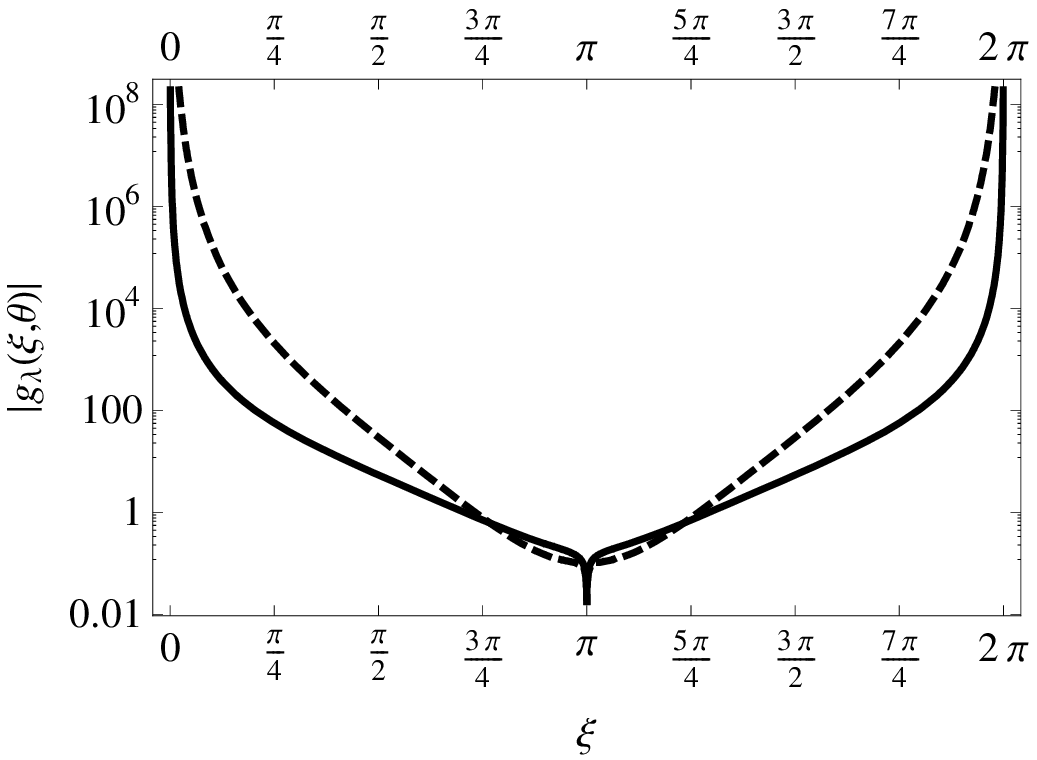}
\caption{Absolute value of the eigenfunctions $f_{\lambda}(\xi,\theta)$ (left panel) and $g_{\lambda}(\xi,\theta)$ (right panel). The black thick line is the eigenfunction for $\lambda=1$, $\theta=1$ and the black dashed line for $\lambda=2$, $\theta=1$.}\label{fig:EF}
\end{center}
\end{figure}

\begin{figure}[!ht]
 \begin{center}
 \includegraphics[width=7.2cm]{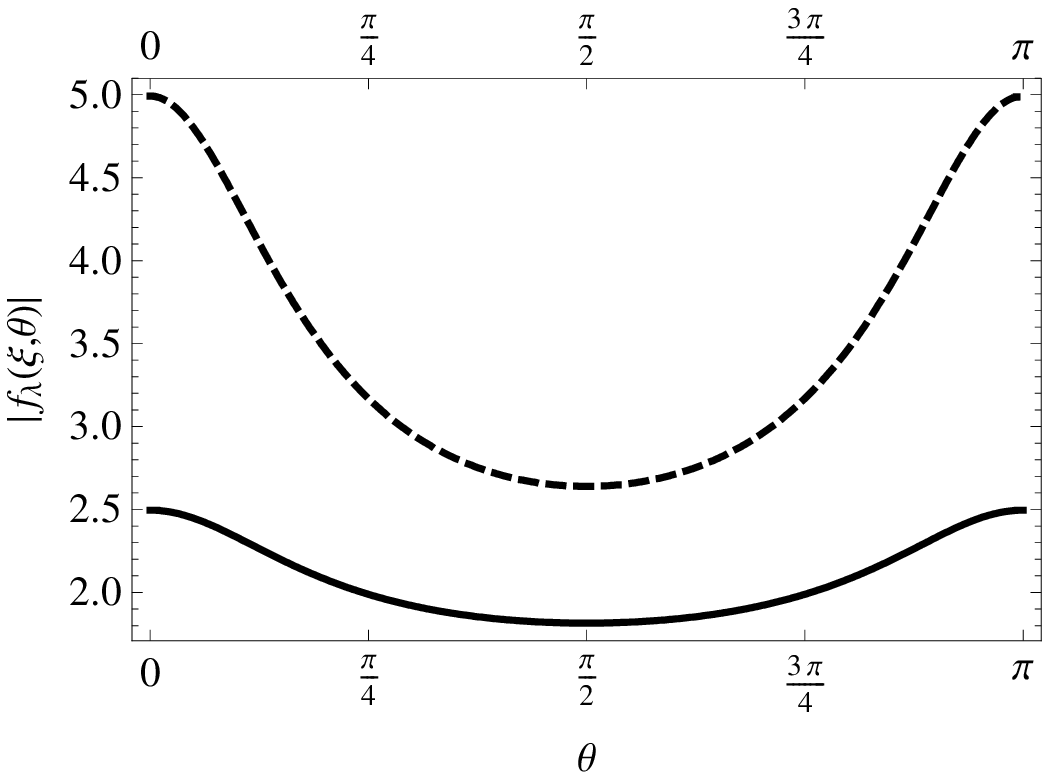}\quad\includegraphics[width=7.2cm]{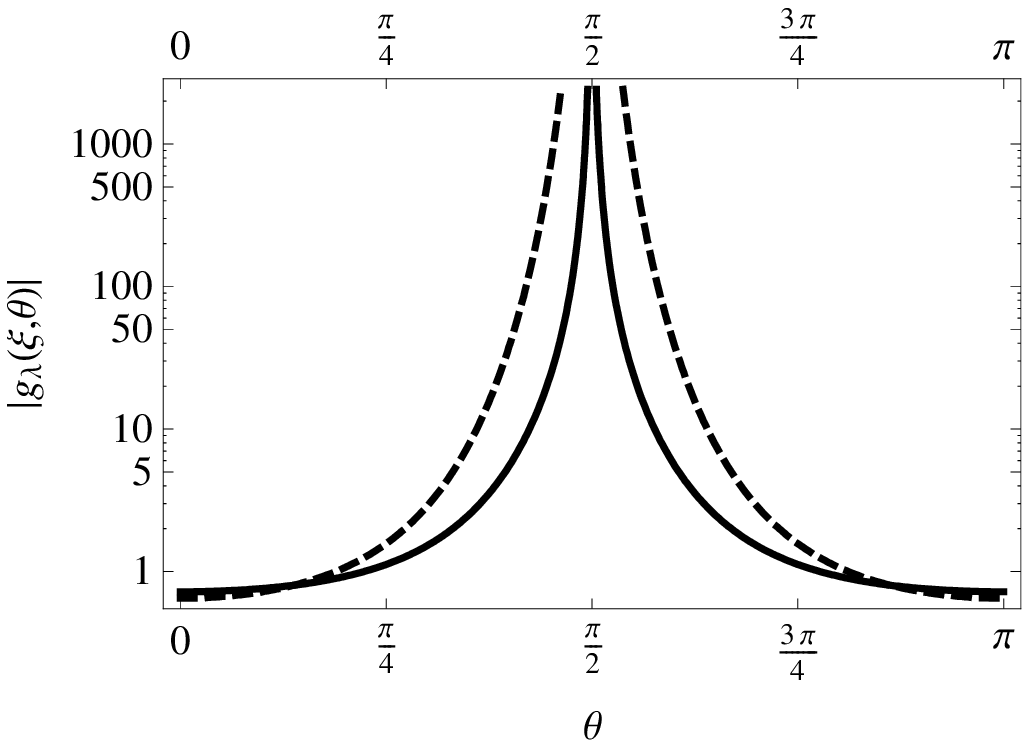}
\caption{Absolute value of the eigenfunctions $f_{\lambda}(\xi,\theta)$ (left panel) and $g_{\lambda}(\xi,\theta)$ (right panel). The black thick line is the eigenfunction for $\lambda=1$, $\xi=2$ and the black dashed line for $\lambda=2$, $\xi=2$.}\label{fig:EFth}
\end{center}
\end{figure}

\subsection{Quantization: 2. Possibility}\label{sec:Schrquantization}
In order to quantize the triad components $\hat p^I\,_i$ we replaced the configuration variables $Q_I$ with $\sin(\delta _IQ_I)/\delta_I$. The question we may ask is to what extend this substitution changes the eigenfunctions. We define the symmetrized operator $\hat A_2$ quantized without the substition of $Q_I$ as
$$\hat A_2=-i\cos\theta\frac{\partial}{\partial \xi}+i\frac{\sin\theta}{\xi}\frac{\partial}{\partial\theta}+\frac{i}{2}\frac{\cos\theta}{\xi}.$$
The solution to the eigenvalue problem $\hat A_2f_{\lambda}(\xi,\theta)=\lambda f_{\lambda}(\xi,\theta)$ is given by
$$f_{\lambda}(\xi,\theta)=\exp\left(i\lambda\xi\cos\theta\right)\sqrt{\xi}\Gamma(\log(\xi\sin\theta)),$$
We see that the eigenfunctions are not almost periodic in $\xi$. However we can choose the function $\Gamma$ such that $\sqrt{\xi}$ disappears, i.e. we set
$$\Gamma=\mathcal{N}_1\exp\left(-\frac{1}{2}\log(\xi\sin\theta)\right),$$
where $\mathcal{N}_1$ is a constant, such that the eigenfunctions to $\hat A_2$ are given by
\be f_{\lambda}(\xi,\theta)=\mathcal{N}_1 \frac{\exp(i\lambda\xi\cos\theta)}{\sqrt{\sin\theta}}.\ee
 The above eigenfunction is almost periodic in $\xi$ but fails to be normalizable on $L^2(\bar\mathds{R}_B)\otimes L^2(U(1))$. As in the preceding section the spectrum of $\hat A_2$ is thus continuous. Note that the eigenfunction is constant in the non-diagonal limit $\theta\rightarrow\pi/2$.

Similarly the eigenfunctions of the symmetrized operator
$$\hat B_2=-i\sin\theta\frac{\partial}{\partial \xi}-i\frac{\cos\theta}{\xi}\frac{\partial}{\partial\theta}+\frac{i}{2}\frac{\sin\theta}{\xi}$$
are given by
\be g_{\lambda}(\xi,\theta)=\mathcal{N}_2\frac{\exp\left(-i\lambda\xi\sin\theta\right)}{\sqrt{\cos\theta}}.\ee
The diagonal limit $\theta\rightarrow0$ of $g_{\lambda}$ is just a constant function such that $\hat p^2\,_3$, i.e. the expectation value $\langle \hat p^2\,_2\rangle$ measures the 'diagonality' of the torus and $\langle \hat p^2\,_3\rangle$ its departure. Once again, the eigenfunctions $g_{\lambda}$ fail to be normalizable on the Hilbert space such that the spectrum of $\hat B_2$ is continuous. Also note that contrary to the first method the commutator between both operators vanishes:
$$[\hat A_2,\hat B_2]=0.$$

\subsection{Volume Operator}\label{sec:kinop}
The classical expression for the volume of $V$ is given by
\begin{equation*}
\mathbf{\mathcal{V}}(V)=\int_V\sqrt{\left|\frac{1}{6}\epsilon_{abc}\epsilon_{ijk}E^{ai}E^{bj}E^{ck}\right|}d^3x.
\end{equation*}
Inserting the definition of the homogeneous densitized triad \eref{defofp} we get:
\begin{equation}\label{volume}
\mathbf{\mathcal{V}}(V)=\sqrt{\mathfrak{k}\left|p^1\,_1(p^2\,_2p^3\,_3-p^2\,_3p^3\,_2)\right|}
\end{equation}
The factor $\mathfrak{k}$ depends on the specific form of the torus and is equal to one if the torus is cubic such that we recover the usual expression in this limit (see e.g. Eq.~(4.5) in \cite{Chiou:07}). Using the classical solution of the Gauss constraint we get the following expression for the physical volume of the torus:
$$\mathbf{\mathcal{V}}(V)=\sqrt{\mathfrak{k}\left|p^1\,_1\left[p^2\,_2p^3\,_3-p^2\,_3\frac{\phi_2\,^2p^2\,_3-\phi_2\,^3p^2\,_2}{\phi_3\,^3}\right]\right|}$$
or in terms of the new variables
\begin{eqnarray}\label{volumeGauss}
\mathbf{\mathcal{V}}(V)&=&\sqrt{\mathfrak{k}\left|\frac{P^1}{Q_2Q_3}\right|}\times\Bigl|\left((P_{\Theta})^2-P^2P^3Q_2Q_3\right)\cos\Theta+\nonumber\\
&&\quad\quad\quad\quad\quad\quad\quad+P_{\Theta}(P^2Q_2+P^3Q_3)\sin\Theta\Bigr|^{1/2}.
\end{eqnarray}

\subsubsection{Quantization of the Volume Operator according to 1. Method}\label{sec:loopvolop}
To perform a quantization of the volume operator we insert the definitions \eref{operators} into $\hat \mathbf{\mathcal{V}}(V)$. Despite the fact that we know the eigenfunctions of the operators $\hat p^I\,_i$ it is not straightforward to give the eigenfunctions of the volume operator $\hat\mathcal{V}$ because, as explained in \Sref{sec:holonomyfluxalgebra}, they do not necessarily commute. Thus the difficult task is to determine the spectrum of the operator
\ba\label{volop}\hat \mathfrak{V}:=\hat p^2\,_2\hat p^3\,_3-\hat p^3\,_2\hat p^2\,_3&=&\frac{\cos\Theta}{\sin Q_2\sin Q_3}\frac{\partial^2}{\partial\Theta^2}-\cos\Theta\frac{\partial^2}{\partial Q_2\partial Q_3}\nonumber\\ &&+\frac{\sin\Theta}{\sin Q_3}\frac{\partial^2}{\partial\Theta\partial Q_2}+\frac{\sin\Theta}{\sin Q_2}\frac{\partial^2}{\partial\Theta\partial Q_3}.\ea
However, this operator is not symmetric on $\HS$. Let us define the symmetric operator 
$$\hat \mathfrak{V}^S:=\frac{1}{2}\left(\hat \mathfrak{V}^++\hat\mathfrak{V}\right).$$
A calculation shows that $\hat\mathfrak{V}^S$ is given by
$$\hat\mathfrak{V}^S=\hat\mathfrak{V}+\frac{1}{2}\left(-\frac{\cos\Theta}{\sin Q_2\sin Q_3}-2\frac{\sin \Theta}{\sin Q_2\sin Q_3}\frac{\partial}{\partial\Theta}+\frac{\cos\Theta}{\sin Q_3}\frac{\partial}{\partial Q_2}+\frac{\cos\Theta}{\sin Q_2}\frac{\partial}{\partial Q_3}\right)$$
This operator is rather complicated and no analytic solutions to the eigenvalue problem could be found.

\subsubsection{Quantization of the Volume Operator according to 2. Method}\label{sec:Schrvolop}
In this subsection we consider the quantization of $\mathcal{V}$ as described in \Sref{sec:Schrquantization} where the commutator between $\hat p^I\,_i$ and $\hat p^J\,_j$ vanishes. This fact simplifies dramatically the analysis because the (generalized) eigenvalue problem can now be written in terms of products and sums of the eigenfunctions of the $\hat p^I\,_i$. Let us define
$$T_{\lambda_1,\lambda_{22},\lambda_{23},\gamma_{33},\gamma_{32}}:=\mathcal{N}_{\lambda_1}\otimes(f_{\lambda_{22}}g_{\lambda_{23}})\otimes (f'_{\gamma_{33}}g'_{\gamma_{32}}),$$
where $f_{\gamma}(Q_2,\theta_1)$, $g_{\gamma}(Q_2,\theta_1)$, $f'_{\gamma}(Q_3,\theta_2)$ and $g'_{\gamma}(Q_3,\theta_2)$ are the (generalized) eigenfunctions of $\hat p^2\,_2$, $\hat p^2\,_3$, $\hat p^3\,_3$ and $\hat p^3\,_2$ respectively given in \Sref{sec:Schrquantization}. Furthermore we denoted the eigenfunctions of $\hat p^1\,_1$ by $\mathcal{N}_{\lambda_1}:=\langle Q_1|\lambda_1\rangle$. Since we have
$$(f_{\lambda_{22}}g_{\lambda_{23}})(f'_{\gamma_{33}}g'_{\gamma_{32}})\propto\frac{\exp\left(iQ_2(\lambda_{22}\cos\theta_1-\lambda_{23}\sin\theta_1)\right)}{\sqrt{\sin\theta_1\cos\theta_1}}\frac{\exp\left(iQ_3(\gamma_{33}\cos\theta_2-\gamma_{32}\sin\theta_2)\right)}{\sqrt{\sin\theta_2\cos\theta_2}}$$
we see that $T_{\lambda_1,\lambda_{22},\lambda_{23},\gamma_{33},\gamma_{32}}$ is not normalizable in $\mathcal{H}^S$. The generalized eigenvalue problem is thus given by
\ba\hat\mathcal{V}T_{\lambda_1,\lambda_{22},\lambda_{23},\gamma_{33},\gamma_{32}}[\varphi]=T_{\lambda_1,\lambda_{22},\lambda_{23},\gamma_{33},\gamma_{32}}[\hat\mathcal{V}\varphi]\nonumber\\
\quad\quad=\gamma^{3/2}\lpl^3\sqrt{k|\lambda_1(\lambda_{22}\gamma_{33}-\lambda_{23}\gamma_{32})|}T_{\lambda_1,\lambda_{22},\lambda_{23},\gamma_{33},\gamma_{32}}[\varphi]\ea
for $\varphi\in\mathcal{D}(\hat\mathcal{V})$.

\subsection{Quantum Gauss Constraint}\label{sec:Gaussconstraint}
In \Sref{sec:closedmodels} we computed the classical Gauss constraint for a Bianchi type I model. In the open case the elementary variables can always be diagonalized such that both the diffeomorphism and Gauss constraints are automatically satisfied. In the closed model this is not the case anymore so that a quantization of the constraints is mandatory. Since in Bianchi type I models the diffeomorphism constraint is proportional to the Gauss constraint we only need to quantize and solve the latter. However, contrary to the diffeomorphism constraint the Gauss constraint can be quantized infinitesimally.

A gauge transformation of an $su(2)$-connection is given by
\ben A\mapsto A'=\lambda^{-1}A\lambda+\lambda^{-1}d\lambda\een
where $\lambda:\Sigma\mapsto SU(2)$. Infinitesimally we can write this equation as
\ben A_a^i\mapsto A^{'i}_a=A_a^i+\partial_a\epsilon^i+\epsilon^i\,_{jk}\epsilon^jA_a^k+\mathcal{O}(\epsilon^2).\een
The classical Gauss constraint ensuring $SU(2)$-invariance is given by
\ben G(\Lambda)=-\int_{\tor}d^3xE^a_jD_a\Lambda^j\een
where $D_a\Lambda^j=\partial_a\Lambda^j+\epsilon^j\,_{kl}A_a^k\Lambda^l$ is the covariant derivative of the smearing field $\Lambda^j$. The infinitesimal quantization of this expression yields an operator containing a sum of right and left invariant vector fields over all vertices and edges of a given graph $\alpha$. This operator is essentially self-adjoint and can, by Stone's theorem, be exponentiated to a unitary operator $U_{\phi}$ defining a strongly continuous one-parameter group in $\phi$. Usually, in order to find the kernel of the Gauss constraint operator one restrict the scalar product on $\mathcal{H}_{\mathrm{aux}}$ to the gauge-invariant scalar product on $\mathcal{H}^G_{\mathrm{inv}}$. This Hilbert space is a true subspace of $\mathcal{H}_{\mathrm{aux}}$ since zero is in the discrete part of the spectrum of the Gauss constraint operator.

We saw in \Sref{sec:cantransf} that thanks to the symmetry reduction two of the Gauss constraints are automatically satisfied. While the nonvanishing Gauss constraint \eref{G1} is still a complicated function in $\phi_I\,^i$ and $p^J\,_j$ it simplifies to \Eref{G1inPQ} after the canonical transformation. A quantization of this expression is then given by
\ben \hat G_1=\hat P_{\theta_1}-\hat P_{\theta_2}.\een
Since the eigenstates of the momentum operators $\hat P_{\theta_{\alpha}}$ are the strict periodic functions satisfying \Eref{eigenvaluesofP} the action of the Gauss constraint on $\ket$ is given by
\ben \hat G_1\ket=\frac{\gamma\lpl^2}{2}(k_1-k_2)\ket\een
which vanishes if
\ben k_1=k_2.\een
We can thus introduce a new variable $\Theta:=\theta_1+\theta_2$ such that the algebra $\mathcal{A}_S$ given by \Eref{gofQP} reduces to the invariant algebra $\mathcal{A}_S^{\mathrm{inv}}$ generated by the functions
\ba\label{ginv} g(Q_1,Q_2,Q_3,\Theta)&=&\sum_{\lambda_1,\lambda_2,\lambda_3,k}\xi_{\lambda_1,\lambda_2,\lambda_3,k}\times \nonumber\\ &&\times\exp\left(\frac{1}{2}i\lambda_1Q_1+\frac{1}{2}i\lambda_2Q_2+\frac{1}{2}i\lambda_3Q_3+ik\Theta\right).\ea
A Cauchy completion leads to the invariant Hilbert space $\mathcal{H}^S_{\mathrm{inv}}=\mathcal{H}_B^{\otimes 3}\times\mathcal{H}_{S^1}$. A comparison with $\HS$ shows that we 'lost' one Hilbert space $\mathcal{H}_{S^1}$ by solving the quantum Gauss constraint. Furthermore, instead of two momentum operators conjugated to $\theta_1$ and $\theta_2$ we have just one momentum operator conjugated to $\Theta$ defined by
$$\hat P_{\Theta}=-i\gamma\lpl^2\frac{\partial}{\partial\Theta}.$$
The eigenstates of all momentum operators are given by
$$|\vec\mu,k\rangle:=|\mu_1,\mu_2,\mu_3,k\rangle,$$
where $k\in\mathds{Z}$ defines the representation of $U(1)$.

\section{Conclusions and Outlook}\label{sec:conclusions}
In this paper we studied how a torus universe affects the results of LQC. To do so we first introduced the most general tori using Thurston's theorem and found that six Teichm\"uller parameters are needed. We construted a metric describing a flat space but respecting the periodicity of the covering group used to construct the torus and used it to derive a gravitational Hamiltonian. We studied the dynamics of a torus universe driven by a homogeneous scalar field by numerically solving the full Hamiltonian and saw that its form only remains cubic if all off-diagonal terms vanish. The Ashtekar connection and the densitized triad for a torus were then derived for both the most general and a slighty simplified torus. The reason for this simplification was that a simple solution to the Gauss constraint could be given. We also derived the Hamiltonian constraint in these new variables and showed that it reduces to the standard constraint of isotropic LQC in case of a cubical torus.

The passage to the quantum theory required a canonical transformation so as to be able to write the holonomies as a product of strictly and almost periodic functions. A Cauchy completion then led to a Hilbert space given by square integrable functions over both $\mathds{R}_B$ and $U(1)$. However the drawback of the canonical transformation is a much more complicated expression for the components of the densitized triad containing both the momentum and the configuration variables. Following the standard procedure of LQC we substituted these configuration variables with the sine thereof and were able to solve the eigenvalue problem analytically. Surprisingly it turned out that the (generalized) eigenfunctions of the triad operators do not lie in the Hilbert space, i.e. the spectrum is continuous. On the other hand we were also able to find almost periodic solutions to the eigenvalue problem of the triad operators without performing the substitution just described, but once again these eigenfunctions do not lie in the Hilbert space. The reason why both ways lead to a continuous spectrum is the non-cubical form of the torus, for if we set the angles $\theta_{1,2}=0$ in \Eref{cantransf} the triads correspond to the ones obtained in isotropic models. Furthermore we were able to find the spectrum of the volume operator for the second case because, contrary to the first case, it is a product of commutating triad operators.

Although we gave a couple of numerical solutions to the classical Hamiltonian we didn't consider its quantization. The constraint describing quantum dynamics of a torus is given by inserting the holonomies \eref{holinQP} into Thiemann's expression for the quantum Hamiltonian \cite{Thiemann:98}
\ben\hat C_{\mathrm{grav}}\propto \epsilon^{ijk}\mathrm{tr}\left(h_i^{(\,^0\lambda_i)}h_j^{(\,^0\lambda_j)}(h_i^{(\,^0\lambda_i)})^{-1}(h_j^{(\,^0\lambda_j)})^{-1}h_k^{(\,^0\lambda_k)}[(h_k^{(\,^0\lambda_k)})^{-1},\hat V]\right). \een
Contrary to LQG and LQC we saw that the spectrum of the volume operator of a torus is continuous. It would thus be very interesting to know how far $\hat C_{\mathrm{grav}}$ departs from the usual difference operator of LQC. Furthermore, whether a quantization of the torus a la LQG removes the Big Band singularity needs also to be addressed, especially since we saw that many characteristics of both LQG and isotropic LQC are not present in this particular topology.

In this work we only considered the simplest closed flat topology but there are many other closed topologies. As we saw there are eight geometries admitting at least one compact quotient. For example there are six different compact quotients with covering $\mathds{E}^3$, namely $\mathds{T}^3$, $\mathds{T}^3/\mathds{Z}_2$, $\mathds{T}^3/\mathds{Z}_3$, $\mathds{T}^3/\mathds{Z}_4$, $\mathds{T}^3/\mathds{Z}_6$ and a space where all generators are screw motions with rotation angle $\pi/2$. It would be interesting to know how these discrete groups $\mathds{Z}_{2,3,4,6}$ affect the results of this work, especially since the last five spaces are inhomogeneous (observer dependent) \cite{Fagundes:92}.

\ack
I would like to thank Martin Bojowald, Frank Steiner and Jan Eric Str\"ang for many useful comments and corrections of previous versions of this manuscript.

\appendix

\section{Fundamental Domain of the Torus}\label{sec:fundamentaldomain}
In two dimensions the upper half-plane $H$ is the set of complex numbers $H=\{x+iy\,|\,y>0;\,x,y\in\mathds{R}\}$. When endowed with the Poincar\'e metric \newline $ds^2=(dx^2+dy^2)/y^2$ this half-plane is called the Poincar\'e upper half-plane and is a two-dimensional hyperbolic geometry. The special linear group $SL(2,\mathds{R})$ acts on $H$ by linear fractional transformations $z\mapsto (az+b)/(cz+d)$, $a,b,c,d\in\mathds{R}$, and is an isometry group of $H$ since it leaves the Poincar\'e metric invariant . The modular group $SL(2,\mathds{Z})\subset SL(2,\mathds{R})$ defines a fundamental domain by means of the quotient space $H/SL(2,\mathds{Z})$. This fundamental domain parametrizes inequivalent families of 2-tori and can thus be identified as the configuration space of the two-dimensional torus. Since we consider a three-dimensional torus with six independent Teichm\"uller parameters (see \Eref{Teichmuellervectors}) we need a generalization of the Poincar\'e upper half-plane \cite{Gordon:87,Terras:88} to a six-dimensional upper half-space.
\begin{Def}
A fundamental domain $D$ for $SL(3,\mathds{Z})$ is a subset of the space \newline $\mathscr{P}_3:=\{A\in Mn(3,\mathds{R})\,|\,A^T=A,\;A\;\mathrm{positive} \; \mathrm{definite}\}$ which is described by the quotient space $\mathscr{P}_3/SL(3,\mathds{Z})$. In other words, if both $A\in\mathscr{P}_3$ and $A[g]:=g^TAg$, $g\in SL(3,\mathds{Z})$, are in the fundamental domain then either $A$ and $A[g]$ are on the boundary of the fundamental domain or $g=id$.
\end{Def}
Since $\mathscr{P}_3$ is a subspace of the six-dimensional Euclidean space (there are six independent matrix elements for $A\in\mathscr{P}_3$), the generalization of the Poincar\'e upper half-plane is now a six-dimensional upper half-space $U^6:=\{(a_1,\ldots,a_6)\in \mathds{E}^6\,|\,a_6>0\}$ upon which the group $SL(3,\mathds{R})$ acts transitively. To identify $\mathscr{P}_3$ with an upper half-space we introduce the Iwasawa coordinates such that $\forall \, A\in\mathscr{P}_3$ there is the unique decomposition:
\begin{equation*}
A=\left(
\begin{array}{ccc}
y_1 & 0&0\\
0&y_2&0\\
0&0&y_3
\end{array}\right)\left(
\begin{array}{ccc}
1&x_1&x_2\\
0&1&x_3\\
0&0&1
\end{array}\right),
\end{equation*}
with $x_i,y_j\in\mathds{R}$ with $\prod y_i=1$. The geometry of the upper half-space $U^6$ is given by the $GL(3,\mathds{R})$-invariant line element 
\begin{eqnarray}\label{dsuhs}
ds^2=\mathrm{tr}((A^{-1}dA)^2)=&&\frac{dy_1\,^2}{y_1\,^2}+\frac{dy_2\,^2}{y_2\,^2}+\frac{dy_3\,^2}{y_3\,^2}.
\end{eqnarray}
Note that the Ricci scalar of the metric \eref{dsuhs} is constant and negative.

In order to give a parametrization of the fundamental region we use Minkowski's reduction theory \cite{Minkowski:05}, which tells us that for a metric $h_{i,j}$ the following inequalities must be satisfied:
\begin{eqnarray*}
h_{i,i}&\leq&h_{i+1,i+1},\quad i=1,2,3\nonumber\\
h_{i,j}&\leq&\frac{1}{2}h_{i,i},\quad i<j.
\end{eqnarray*}
Since the metric \eref{metrich} is invariant under the map $a_3\,^3\rightarrow -a_3\,^3$ we can define the upper half-space as $U^6=\{(a_1\,^1,a_2\,^1,a_2\,^2,a_3\,^1,a_3\,^2,a_3\,^3)\in\mathds{E}^6\,|\,a_3\,^3>0\}$, where we have identified the element $a_6$ with $a_3\,^3$. In our parametrization~\eref{metrich} we therefore obtain the fundamental domain $D$ delimited by the inequalities:
\begin{eqnarray*}
(a_1\,^1)^2\leq (a_2\,^1)^2+(a_2\,^2)^2\leq (a_3\,^1)^2+(a_3\,^2)^2+(a_3\,^3)^2\nonumber \\
a_2\,^1\leq\frac{1}{2}a_1\,^1\nonumber \\
a_3\,^1\leq\frac{1}{2}a_1\,^1 \nonumber \\
a_2\,^1a_3\,^1+a_2\,^2a_3\,^2\leq \frac{1}{2}\left((a_2\,^1)^2+(a_2\,^2)^2\right)
\end{eqnarray*}

The first inequality tells us that the length of the generators of the torus are ordered: $\|a_1\|\leq\|a_2\|\leq\|a_3\|$. However, starting with such an ordered triplet does not necessarily imply that the order is preserved by dynamics. Thus we think that it may be more appropriate to choose the equivalent representation of the configuration space given by $\mathcal{C}=\mathds{R}^6$. Otherwise, we would have to rotate the coordinate system every time the torus leaves the fundamental domain. Note that similar results have also been obtained in M-theory, where one considers a compactification of the extra dimensions on $\mathds{T}^n$ (see e.g. \cite{McGuigan:90,McGuigan:03}). However, the situation is different in string theory where one really integrates only over the fundamental domain, e.g. $Z(\mathds{T}^n)=\int_D d\bm{\tau} Z(\bm{\tau})$.

\section{The Torus Universe in Iwasawa Coordinates}\label{sec:TorusIwasawa}
In this appendix we use a parametrization of the torus using the Iwasawa coordinates which are more apt to describe the asymptotic behavior of the metric \cite{Damour:03}. It is important to understand the role of the off-diagonal terms in the metric \eref{metrich} and to know what happens near the singularity and at late times. The metric can be decomposed as follows:
\begin{equation}\label{metricI}
h=\mathcal{N}^T\mathcal{D}^2\mathcal{N},
\end{equation}
where
\begin{equation*}
\mathcal{D}=\left(
\begin{array}{ccc}
e^{-z_1} & 0&0\\
0&e^{-z_2}&0\\
0&0&e^{-z_3}
\end{array}\right), \quad
\mathcal{N}=\left(
\begin{array}{ccc}
1 & n_1&n_2\\
0&1&n_3\\
0&0&1
\end{array}\right).
\end{equation*}
An easy calculation shows that \Eref{metricI} can be transformed into \Eref{metrich} with $n_1=a_2\,^1/a_1\,^1$, $n_2=a_3\,^1/a_1\,^1$, $n_3=a_3\,^2/a_2\,^2$, $z_i=-\ln a_i\,^i$ (no summation)\footnote{For simplicity we assume that all diagonal scale factors $a_i\,^i$ are strictly positive. However the nondiagonal scale factors can be negative or zero.}. The analogue to \Eref{ds2Teich} is now given by
\begin{equation}\label{habI}
h_{ab}=\sum_{i=1}^3e^{-2z_i}\mathcal{N}_a\,^i\mathcal{N}_b\,^i.
\end{equation}
The Iwasawa decomposition can also be viewed as the Gram-Schmidt orthogonalization of the forms $dx^a$:
\begin{equation*}
h_{ab}dx^a\otimes dx^b=\sum_{i=1}^3e^{-2z_i}\theta^i\otimes \theta^i,
\end{equation*}
where the coframes $\theta^i$ are given by
\begin{equation*}
\theta^i=\mathcal{N}_a\,^idx^a.
\end{equation*}
Analogously, the frames $e_i$ dual to the coframes $\theta^i$ are given by the inverse of $\mathcal{N}_a\,^i$:
\begin{equation*}
e_i=\mathcal{N}^a\,_i\frac{\partial}{\partial x^a}.
\end{equation*}
Since the determinant of the matrix $\mathcal{N}$ is equal to one the basis given by the coframe is orthonormal. Note that this is different from the construction in \Sref{sec:torusuniverse}.

In order to determine the asymptotic behavior of the off-diagonal terms we follow the analysis in \cite{Damour:03}. The metric $h$ being symmetric, we automatically know that its eigenvalues are real. We call these eigenvalues $t^{2\alpha_i}$, $1\leq i \leq3$ and $\alpha_1<\alpha_2<\alpha_3$, in analogy to the diagonal Kasner solution (see \Sref{sec:openmodels}) and construct a metric $h_K(t)=\mathrm{diag}(t^{2\alpha_i})$ by means of a constant matrix $L$
\begin{equation*}
h(t)=L^Th_K(t)L, \quad L=\left(
\begin{array}{ccc}
l_1 & l_2 & l_3\\
m_1 & m_2 & m_3\\
r_1 & r_2 & r_3
\end{array}\right).
\end{equation*}
With these relations we can obtain the evolution of the Iwasawa variables. For example, we have
\begin{equation*}
n_1(t)=\frac{t^{2\alpha_1}l_1l_2+t^{2\alpha_2}m_1m_2+t^{2\alpha_3}r_1r_2}{t^{2\alpha_1}l_1^2+t^{2\alpha_2}m_1^2+t^{2\alpha_3}r_1^2}.
\end{equation*}
In \cite{Damour:03} it was shown that the asymptotic behavior $t\rightarrow 0^+$ of the off-diagonal terms of the Iwasawa variables is given by
\begin{equation*}
n_1\rightarrow \frac{l_2}{l_1},\quad n_2\rightarrow \frac{l_3}{l_1},\quad n_3\rightarrow\frac{l_1m_3-l_3m_1}{l_1m_2-l_2m_1}, \quad (t\rightarrow 0^+),
\end{equation*}
which means that the off-diagonal terms freeze in as we approach the singularity. We can also calculate the other limit $t\rightarrow\infty$ and obtain e.g.
\begin{equation*}
n_1\rightarrow \frac{r_2}{r_1},\quad n_2\rightarrow\frac{r_3}{r_1},\quad n_3\rightarrow\frac{m_1r_3-m_3r_1}{m_1r_2-m_2r_1},\quad(t\rightarrow \infty).
\end{equation*}
We have checked this result numerically, which can be seen in \Fref{fig:torus_sim} where the gray line $(a_2\,^1)$ converges  for $t\rightarrow\infty$ toward the solid line ($a_1\,^1)$, i.e. $n_1\rightarrow \mathrm{const}$. However, the limit $t\rightarrow0^+$ could not be checked due to the numerical instability of the solutions when approaching the singularity.

\section*{References}


\begin{thebibliography}{99}
\bibitem{Komatsu:08} Komatsu E {\it et al}, {\it Five-Year Wilkinson Microwave Anisotropy Probe (WMAP) Observations: Cosmological Interpretation}, 2009 {\it Astrophys.J.Suppl.} {\bf 180} 330
\bibitem{Aurich:08} Aurich R, Janzer H S, Lustig S and Steiner F, {\it Do we live in a 'small universe'}, 2008 {\it Class. Quantum Grav.} {\bf 25} 125006
\bibitem{Aurich:08:2} Aurich R, {\it A spatial correlation analysis for a toroidal universe}, 2008 {\it Class. Qautnum Grav.} {\bf 25} 225017
\bibitem{Aurich:09} Aurich R, Lustig S and Steiner F, {\it Hot pixel contamination in the CMB correlation function}, 2009 {\texttt astro-ph.CO/0903.3133}
\bibitem{Aurich:05} Aurich R, Lustig S and Steiner F, {\it CMB Anisotropy of the Poincar\'e dodecahedron}, 2005 {\it  Class. Quantum Grav.} {\bf 22} 2061
\bibitem{Caillerie:07} Caillerie S {\it et al}, {\it A New Analysis of the Poincar\'e Dodecahedral Space Model}, 2007 {\it A \& A} {\bf 476} 691C
\bibitem{Aurich:05:2} Aurich R, Lustig S and Steiner F, {\it CMB Anisotropy of Spherical Spaces}, 2005 {\it  Class. Quantum Grav.} {\bf 22} 3443
\bibitem{Bojowald:08} Bojowald M, {\it Loop Quantum Cosmology}, 2008 {\it Living Re. Relativity} {\bf 11} 4
\bibitem{Bojowald:00} Bojowald M and Kastrup H A, {\it Symmetry Reduction for Quantized Diffeomorphism-invariant Theories of Connections}, 2000 {\it Class. Quantum Grav.} {\bf 17} 3009
\bibitem{Bojowald:00:2} Bojowald M, {\it Loop Quantum Cosmology I: Kinematics}, 2000 {\it Class. Quantum Grav.} {\bf 17} 1489
\bibitem{Bojowald:00:3} Bojowald M, {\it Loop Quantum Cosmology II: Volume Operator}, 2000 {\it Class. Quantum Grav.} {\bf 17} 1509
\bibitem{Bojowald:02} Bojowald M, {\it Isotropic Loop Quantum Cosmology}, 2002 {\it Class. Quantum Grav.} {\bf 19} 2717
\bibitem{Ashtekar:03} Ashtekar A, Bojowald M and Lewandowski J, {\it Mathematical Structure of Loop Quantum Cosmology}, 2003 {\it Adv. Theor. Math. Phys.} {\bf 7} 233
\bibitem{Bojowald:03} Bojowald M, {\it Homogeneous Loop Quantum Cosmology}, 2003 {\it Class. Quantum Grav.} {\bf 20} 2595
\bibitem{Ashtekar:06} Ashtekar A, Pawlowski T and Singh P, {\it Quantum Nature of the Big Bang: An Analytical and Numerical Investigation}, 2006 {\it Phys. Rev. D} {\bf 73} 124038
\bibitem{Ashtekar:06:2} Ashtekar A, Pawlowski T and Singh P, {\it Quantum Nature of the Big Bang: Improved Dynamics}, 2006 {\it Phys.Rev. D} {\bf 74} 084003
\bibitem{Ashtekar:04} Ashtekar A and Lewandowski J, {\it Background Independent Quantum Gravity: A Status Report}, 2004 {\it Class. Quantum Grav.} {\bf 21} R53
\bibitem{Rovelli:04} Rovelli C, {\it Quantum Gravity}, 2004 (Cambridge: Cambridge University Press)
\bibitem{Thiemann:07} Thiemann T, {\it Modern Canonical Quantum General Relativity}, 2007 (Cambridge: Cambridge University Press)
\bibitem{thooft:79} 't Hooft G, {\it A property of electric and magnetic flux in non-abelian gauge theories}, 1979 {\it Nucl. Phys.} B {\bf 153} 141
\bibitem{Ashtekar:91} Ashtekar A and Samuel J, {\it Bianchi Cosmologies: the Role of Spatial Topology}, 1991 {\it Class. Quantum Grav.} {\bf 8} 2191
\bibitem{Wolf:74} Wolf J A, {\it Spaces of Constant Curvature}, 1974 (Boston: Publish Or Perish)
\bibitem{Koike:94} Koike T, Tanimoto M and Hosoya A, {\it Compact Homogeneous Universes}, 1994 {\it J. Math. Phys.} {\bf 35} 4855
\bibitem{Tanimoto:97} Tanimoto M, Koike T and Hosoya A, {\it Dynamics of Compact Homogeneous Universes}, 1997 {\it J. Math. Phys.} {\bf 38} 350
\bibitem{Tanimoto:97:02} Tanimoto M, Koike T and Hosoya A, {\it Hamiltonian Structures for Compact Homogeneous Universes}, 1997 {\it J. Math. Phys.} {\bf 38} 6560
\bibitem{Yasuno:01} Yasuno K, Koike T and Siino M, {\it Thurston's Geometrization Conjecture and Cosmological Models}, 2001 {\it Class. Quantum Grav.} {\bf 18} 1405
\bibitem{Singer:60} Singer I M, {\it Infinitesimally Homogeneous Spaces}, 1969 {\it Comm. Pure Appl. Math.} {\bf 13} 685
\bibitem{Thurston:97} Thurston W, {\it Three-dimensional geometry and topology}, 1997 Vol. 1. (Princeton: Princeton University Press)
\bibitem{Kobayashi:63} Kobayashi S and Nomizu K, {\it Foundations of Differential Geometry}, volume 1 (John Wiley \& Sons, New York 1963); volume 2 (New York 1969)
\bibitem{Brodbeck:96} Brodbeck O, {\it On Symmetric Gauge Fields for Arbitrary Gauge and Symmetry Groups}, 1996 {\it Helv. Phys. Acta} {\bf 69} 321
\bibitem{Chiou:07} Chiou D-W, {\it Loop Quantum Cosmology in Bianchi Type I Models: Analytical Investigation}, 2007 {\it Phys.Rev.} D{\bf 75} 024029
\bibitem{Ashtekar:06:3} Ashtekar A and Bojowald M, {\it Quantum Geometry and the Schwarzschild Singularity}, 2006 {\it Class. Quantum Grav.} {\bf 23} 391
\bibitem{Thiemann:98:4} Thiemann T, {\it Kinematical Hilbert Spaces for Fermionic and Higgs Quantum Field Theories}, 1998 {\it Class. Quantum Grav.} {\bf 15} 1487
\bibitem{Velhinho:07} Velhinho J M, {\it The Quantum Configuration Space of Loop Quantum Cosmology}, 2007 {\it Class. Quantum Grav.} {\bf 24} 3745
\bibitem{Bratteli:79} Bratelli O and Robinson D W, {\it Operator Algebras and Quantum Statistical Mechanics}, 1979 (New York: Springer)
\bibitem{Bojowald:04} Bojowald M and Swiderski R, {\it The Volume Operator in Spherically Symmetric Quantum Geometry}, 2004 {\it Class. Quantum Grav.} {\bf 21} 4881
\bibitem{Kamke:79} Kamke E, {\it Diffenrentialgleichungen, L\"osungsmethoden und L\"osungen II}, 1979 (Stuttgart: B.G. Teubner)
\bibitem{Shubin:74} Shubin M A, {\it Differential and pseudodifferential operators in spaces of almost periodic functions}, 1974 {\it Math. USSR Sbornik} {\bf 24} 547
\bibitem{Shubin:78} Shubin M A, {\it Almost periodic functions and partial differential operators}, 1978 {\it Russian Math. Surveys} {\bf 33} 1
\bibitem{Bohr:47} Bohr H, {\it Almost Periodic Functions}, 1947 Chelsea Publishing Company
\bibitem{Roberts:66} Roberts J E, {\it The {D}irac bra and ket formalism}, 1966 {\it J. Math. Phys.} {\bf 7} 1097
\bibitem{Roberts:66:2} Roberts J E, {\it Rigged {H}ilbert Space in quantum mechanics}, 1966 {\it Commun. Math. Phys.} {\bf 3} 98
\bibitem{Thiemann:98} Thiemann T, {\it Anomaly-free Formulation of Non-perturbative, Four-dimensional Lorentzian Quantum Gravity}, 1998 {\it Phys. Lett.} B {\bf 380} 257
\bibitem{Fagundes:92} Fagundes H V, {\it Closed Spaces in Cosmology}, 1992 {\it Gen. Rel. Grav.} {\bf 24} 199
\bibitem{Gordon:87} Gordon D, Grenier D and Terras A, {\it Hecke Operators and the Fundamental Domain of $SL(3,\mathds{Z})$}, 1987 {\it Math. Comp.} {\bf 48} 159
\bibitem{Terras:88} Terras A, {\it Harmonic Analysis on Symmetric Spaces and Applications II}, 1988 Vol. II (New York, Springer Verlag)
\bibitem{Minkowski:05} Minkowski H, {\it Diskontinuit\"atsbereich f\"ur arithmetische \"Aquivalenz}, 1905 {\it J. Reine Angew. Math.} {\bf 129} 220
\bibitem{McGuigan:90} McGuigan M, {\it Fundamental Regions of Superspace}, 1990 {\it Phys. Rev.} D {\bf 41} 1844
\bibitem{McGuigan:03} McGuigan M, {\it Three Dimensional Gravity and M-Theory}, 2003 {\texttt hep-th/0312327}
\bibitem{Damour:03} Damour T, Henneaux M and Nicolai H, {\it Cosmological Billards}, 2003 {\it Class. Quantum Grav.} {\bf 20} R145-R200
\end{thebibliography}
\end{document}